# Optical vortex harmonic generation facilitated by photonic spin-orbit entanglement

**Chang Kyun Ha, Eun Mi Kim, Kyoung Jun Moon, and Myeong Soo Kang***

Department of Physics, Korea Advanced Institute of Science and Technology (KAIST)

291 Daehak-ro, Yuseong-gu, Daejeon 34141, Republic of Korea

*mskang@kaist.ac.kr

**Photons can undergo spin-orbit coupling, by which the polarization (spin) and spatial profile (orbit) of the electromagnetic field interact and mix. Strong photonic spin-orbit coupling may reportedly arise from light propagation confined in a small cross-section, where the optical modes feature spin-orbit entanglement. However, while photonic Hamiltonians generally exhibit nonlinearity, the role and implication of spin-orbit entanglement in nonlinear optics have received little attention and are still elusive. Here, we report the first experimental demonstration of nonlinear optical frequency conversion, where spin-orbit entanglement facilitates spin-to-orbit transfer among different optical frequencies. By pumping a multimode optical nanofiber with a spin-polarized Gaussian pump beam, we produce an optical vortex at the third harmonic, which has long been regarded as a forbidden process in isotropic media. Our findings offer a unique and powerful means for efficient optical vortex generation that only incorporates a single Gaussian pump beam, in sharp contrast to any other approaches employing structured pump fields or sophisticatedly designed media. Our work opens up new possibilities of spin-orbit-coupling subwavelength waveguides, inspiring fundamental studies of nonlinear optics involving various types of structured light, as well as paving the way for the realization of hybrid quantum systems comprised of telecom photonic networks and long-lived quantum memories.**

## Introduction

Spin-orbit coupling (SOC)—the interaction and mixing between the spin and orbital degrees of freedom (DOFs) of a particle—arises in diverse disciplines. In individual atoms and molecules, SOC occurs via the orbital motion of spin-1/2 electrons in the nucleus-generating electrostatic potential, contributing to the formation of fine structures [1]. An analogous energy level splitting takes place inside an atomic nucleus through the cooperation of the orbital motion and strong nuclear force for protons and neutrons [2]. In condensed matters, SOC plays a crucial role in the emergence of various fundamental phenomena involving electronic spins, such as the spin Hall effect and anomalous Hall effect [3], Rashba effect [4], antisymmetric exchange [5,6], and spin-momentum locking [7]. These phenomena have been investigated for the generation and control of spin currents [3], spin-orbit torques [8], and magnetic skyrmions [9], as well as the implementation of functional spintronic devices for faster and more efficient information storage and processing [10].

Photons also possess spin and orbital DOFs, which are manifested in the polarization and spatial distribution, respectively, of the electromagnetic field, and can accordingly carry spin angular momentum (SAM) and orbital angular momentum (OAM) [11] like other particles. In particular, the optical beam possessing a nonzero OAM exhibits a helically twisted wavefront and a donut-shaped intensity profile with an intensity null at the beam center, so it is referred to as an optical vortex [12]. Moreover, the two DOFs of light can interact and mix in some circumstances. Such photonic SOC has been exploited primarily to generate and manipulate optical vortices through spin-to-orbit transfer, being at the heart of the remarkable advances in optical communications [13], optical trapping [14], and optical imaging [15,16], as well as the recent advent of novel technologies, such as nonclassical light generation [17,18], complex lattice pattern generation [19], spiral phase contrast imaging with invisible illumination [20], and high-dimensional photonic quantum information processing [21,22].

More recently, it has been pointed out that photonic SOC can arise from simple light confinement in a small cross-sectional area. When light is strongly focused by a high-numerical-aperture lens [23,24] or tightly confined within a subwavelength waveguide [25–27], the longitudinal field component becomes significant, so the light field loses the transversality. Then, the propagation eigenmodes of photons cannot be represented as a simple product of spin and orbital states (i.e., SAM/OAM eigenmode) any longer but appear as spin-orbit-entangled states [25,26], akin to the total angular momentum (TAM) eigenstates in the spin-orbit-coupled atomic fine structure [1]. Since such spin-orbit entanglement (SOE) is a purely geometrical effect, it can occur in a broad range of micro/nano-photonic systems made of any materials, including even isotopic ones (e.g., fused silica glass). While the photonic Hamiltonian given in terms of the electric permittivity of a medium can involve nonlinearity, which is a sharply distinctive feature to the intrinsically linear counterparts of other quantum particles, the role and implication of the confinement-induced SOC and the resulting SOE in nonlinear optics have been rarely studied and thus still elusive.

In this paper, we experimentally reveal that nonlinear optical frequency conversion can accompany spin-to-orbit transfer among different optical frequencies when the nonlinear photonic Hamiltonian is created by spin-orbit-entangled light. By pumping a multimode optical nanofiber [28] that tightly confines optical fields with a circularly polarized Gaussian beam (possessing a spin ±1 but no OAM), we generate an optical vortex having a nonzero OAM at the third harmonic optical frequency. This observation indicates that spin-to-orbit conversion occurs from the pump to the third harmonic in the third-harmonic generation (THG) process, even though the nanofiber is axially symmetric and made entirely of isotropic material (fused silica glass). This unprecedented process, which we term third-harmonic optical vortex generation (TH-OVG), cannot be observed in isotropic bulk media in the paraxial regime (i.e., without SOC) and weakly guiding conventional optical fibers. In reality, THG has long been perceived to be forbidden in such media when pumped with a circularly polarized beam [29–31]. Our full-vectorial nonlinear coupled-mode theory convinces us that the multimode nanofiber exhibiting confinement-induced photonic SOC supports spin-orbit-entangled guided eigenmodes, which plays an essential role in facilitating the TH-OVG process. Our result also offers a new mechanism of all-optical vortex generation, which operates with a single Gaussian pump beam only and relies exclusively on the confinement-induced SOC, not requiring any structured light, specially designed medium, or particular class of material. Our scheme is thus simple and cost-effective while equipped with ultrafast all-optical switching at the same time and should be distinguished from any other approaches that employ sophisticatedly structured components (e.g., spiral phase plates [32], Q plates [33], computer-generated holograms [34], spatial light modulators [35], microcavities [36], on-chip gratings [37], and metasurfaces [38,39]) or nonlinear optical phenomena in structured media [40,41] or with complex pumping configurations using multiple/structured laser beams [42–46]. We focus on harmonic generation here, but our concept can be applied to any nonlinear optical wave mixing processes.

## Principle and theoretical analysis

An optical waveguide with a circular cross-section supports a family of optical vortices as the guided eigenmodes. Each optical vortex mode (OVM), which we denote here by $\mathrm{OV}_{l,m}^{\sigma}$, can be formed by linear superposition of the even and odd hybrid modes with the phase difference of 90° as below [27,47].

$$\begin{aligned} \mathrm{OV}_{\pm l,m}^{\pm} &= \mathrm{HE}_{l+1,m}^{\mathrm{even}} \pm i\,\mathrm{HE}_{l+1,m}^{\mathrm{odd}}, \\ \mathrm{OV}_{\pm l,m}^{\mp} &= \mathrm{EH}_{l-1,m}^{\mathrm{even}} \pm i\,\mathrm{EH}_{l-1,m}^{\mathrm{odd}}, \end{aligned} \tag{1}$$

where $l$ and $\sigma$ represent the canonical OAM and SAM numbers, respectively, in the paraxial regime, and $m$ is the radial mode number. For instance, the circularly polarized Gaussian-like fundamental $\mathrm{HE}_{11}$ mode is designated as the $\mathrm{OV}_{0,1}^{\pm}$ mode, which carries an SAM of $\sigma = \pm 1$ but no OAM ($l = 0$). (See Supplementary Note 1 for more detail) In the paraxial regime, each OVM is a simultaneous OAM and SAM eigenmode, both the OAM and SAM eigenvalues being integers [25,26]. In this case, the OAM and SAM should be individually conserved in any nonlinear optical processes, so their mutual exchange (e.g., spin-to-orbit conversion and vice versa) is prohibited [43]. The OVMs in weakly guiding conventional optical fibers also have these properties approximately. In contrast, when the OVM is confined within a sufficiently small cross-sectional area, it loses the field transversality and exhibits significant SOC associated with the longitudinal field component [25–27], which can be expressed by the normalized field distribution of the OVM as follows.

$$\begin{aligned} \mathbf{e}_{+} &= e^{+ij\phi}\left[ iA_{j,m}(r)e^{-i\phi}\hat{\boldsymbol{\sigma}}_{+} + iB_{j,m}(r)e^{+i\phi}\hat{\boldsymbol{\sigma}}_{-} + C_{j,m}(r)\hat{\mathbf{z}} \right], \\ \mathbf{e}_{-} &= e^{-ij\phi}\left[ iB_{j,m}(r)e^{-i\phi}\hat{\boldsymbol{\sigma}}_{+} + iA_{j,m}(r)e^{+i\phi}\hat{\boldsymbol{\sigma}}_{-} + C_{j,m}(r)\hat{\mathbf{z}} \right], \end{aligned} \tag{2}$$

where $j = l + \sigma$ is the TAM number of the OVM, and $\mathbf{e}_{+}$ and $\mathbf{e}_{-}$ correspond to the even+*i*odd OVM of left-handedness and the even–*i*odd OVM of right-handedness, respectively. $\hat{\boldsymbol{\sigma}}_{\pm} = (\hat{\mathbf{x}} \pm i\hat{\mathbf{y}})/\sqrt{2}$ are the transverse unit vectors along the left-handed ($\hat{\boldsymbol{\sigma}}_{+}$) and right-handed ($\hat{\boldsymbol{\sigma}}_{-}$) circular polarizations, while $\hat{\mathbf{z}}$ is the axial unit vector, and $(r, \phi, z)$ are the cylindrical coordinates. (See Supplementary Note 3 for more detail) Equation (2) indicates that the OVM is no longer the OAM or SAM eigenmode but becomes a spin-orbit-entangled TAM eigenmode.

In this scenario, neither the OAM nor SAM is necessarily preserved in a nonlinear optical process, while the TAM must still be conserved [27]. This feature of spin-orbit-entangled OVMs may allow optical vortex generation through spin-to-orbit-converting nonlinear wave mixing of OVMs.

In this work, we exploit the characteristics of the strongly spin-orbit-entangled OVMs tightly confined in a multimode optical nanofiber to reveal the optical-vortex-creating unconventional nonlinear frequency conversion process. Let us focus on THG pumped by a circularly polarized Gaussian beam having no OAM ($l_p$ = 0) but an SAM of $\sigma_p$ = ±1, as illustrated in Fig. 1(a). Since a single third-harmonic photon is created at the expense of the annihilation of three pump photons in the THG process, the TAM of the third-harmonic photon should be $j_s$ = ±3 because that of each pump photon is $j_p$ = ±1 (Fig. 1(b)). One might think that this THG process can occur naturally and lead to optical vortex creation via spin-to-orbit conversion, as the SAM of the third-harmonic photon can be $\sigma_s$ = ±1 only. However, the THG process is forbidden in isotropic bulk media in the paraxial regime and weakly-guiding conventional optical fibers [29–31], which is related to the fact mentioned earlier that the OAM and SAM should be preserved separately in the nonlinear frequency conversion. In fact, it has long been perceived that THG cannot occur when pumped by a circularly polarized beam, and thus most previous THG experiments employed a linearly polarized pump beam exclusively [29–31]. In sharp contrast, when the circularly polarized pump beam is launched into the Gaussian-like $\mathrm{OV}_{0,1}^{\pm}$ mode of a multimode nanofiber, a third harmonic can be generated efficiently if the phase-matching condition is satisfied. The phase-matching nanofiber diameter [28,48] is so small (717 nm in our case in Fig. 1(c)) that the guided OVMs can exhibit strong SOC, and either the OAM or the SAM does not have to be conserved in the THG process. We will show that, as a consequence of the SOC-induced SOE, third-harmonic photons in another OVM with the TAM of $j_s$ = ±3 can be generated efficiently, which can eventually evolve into a circularly polarized optical vortex with an integer OAM ($l_s$ = ±2) and SAM ($\sigma_s$ = ±1) in the weakly guiding or paraxial regime.

We perform numerical modeling of the OVMs to verify the experimental feasibility of the TH-OVG. The 1550-nm pump in the fundamental $\mathrm{OV}_{0,1}^{\pm}$ mode is phase-matched with the third harmonic in the higher-order OVMs at specific diameters of silica glass optical nanofiber (Fig. 1(c)). Among various third-harmonic OVMs, we concentrate on the $\mathrm{OV}_{\pm2,1}^{\pm}$ mode ($HE_{31}$ hybrid mode) because it has a TAM of ±3 and thus may be generated through TH-OVG. The phase-matching nanofiber diameter is calculated to be 717 nm. Figure 1(d) displays the field profiles of the OVMs and the constituent hybrid modes in the 717-nm-thick nanofiber, where SOC appears as the nonuniformity in the polarization and phase patterns on the beam cross-sections. (See Supplementary Figure S1, which also displays the OVMs of right-handedness) We note that the degeneracy between the $\mathrm{OV}_{\pm2,1}^{\pm}$ and the $\mathrm{OV}_{\pm2,1}^{\mp}$ mode ($EH_{11}$ hybrid mode) is broken significantly, as seen in Fig. 1(c). These two modes are the spin-orbit aligned and anti-aligned OVMs, respectively, of the same OAM number of ±2, and the evident non-degeneracy between them verifies the action of strong SOC.

To reveal and elucidate the role of the SOC-induced SOE in the TH-OVG process, we derive the full-vectorial nonlinear coupled-mode equations that describe the TH-OVG in the OVM basis as below.

$$\frac{da_{\mathrm{s},\pm}}{dz} = i\frac{9}{16}kZ_0\chi_{xxxx}^{(3)}J_{\mathrm{T,OVM}}a_{\mathrm{p},\pm}^{3}e^{i\Delta\beta z}, \tag{3}$$

$$\begin{aligned} J_{\mathrm{T,OVM}} &= \frac{2}{3}\int\Big[\left(\mathbf{e}_{\mathrm{p},+}\cdot\mathbf{e}_{\mathrm{p},+}\right)\left(\mathbf{e}_{\mathrm{p},+}\cdot\mathbf{e}_{\mathrm{s},+}^{*}\right)+\left(\mathbf{e}_{\mathrm{p},-}\cdot\mathbf{e}_{\mathrm{p},-}\right)\left(\mathbf{e}_{\mathrm{p},-}\cdot\mathbf{e}_{\mathrm{s},-}^{*}\right)\Big]dA \\ &\propto \int d\phi\cos\Big[\left(j_{\mathrm{s}}-3j_{\mathrm{p}}\right)\phi\Big], \end{aligned} \tag{4}$$

Here, $J_{\mathrm{T,OVM}}$ is the effective overlap integral that represents the spatial overlap between the pump OVM and the third-harmonic OVM for TH-OVG. *a* is the slowly varying field amplitude of each OVM, and the subscripts p and s stand for the pump and the third harmonic, respectively. (See Supplementary Notes 2 and 3 for more detail) Equation (4) indicates that $J_{\mathrm{T,OVM}}$ is nonzero in the presence of SOC only if the TAM can be conserved, i.e., $j_{\mathrm{s}} = 3j_{\mathrm{p}}$, whereas the requirement for the simultaneous conservation of both OAM and SAM is released. Consequently, the THG in the $\mathrm{OV}_{\pm2,1}^{\pm}$ mode with $j_{\mathrm{s}} = \pm3$ is permitted with a circularly polarized pump beam in the fundamental $\mathrm{OV}_{0,1}^{\pm}$ mode having $j_{\mathrm{p}} = \pm1$. In contrast, $J_{\mathrm{T,OVM}}$ becomes identically zero for the $\mathrm{OV}_{\pm2,1}^{\mp}$ and the $\mathrm{OV}_{0,2}^{\pm}$ modes with $j_{\mathrm{s}} = \pm1$, so THG in those OVMs is forbidden. (See Supplementary Table 1 for the calculated values of overlap integrals for three third-harmonic OVMs, the $\mathrm{OV}_{\pm2,1}^{\pm}$ ($HE_{31}$), the $\mathrm{OV}_{\pm2,1}^{\mp}$ ($EH_{11}$), and the $\mathrm{OV}_{0,2}^{\pm}$ ($HE_{12}$) modes) On the other hand, without SOC, the spin and orbital DOFs of an OVM are separated from each other, as only either $A_{j,m}(r)$ or $B_{j,m}(r)$ is nonzero in Eq. (2). $J_{\mathrm{T,OVM}}$ then takes the form of

$$J_{\mathrm{T,OVM}} \propto \delta_{\sigma_{\mathrm{s}},3\sigma_{\mathrm{p}}} \int d\phi \cos\left[\left(l_{\mathrm{s}} - 3l_{\mathrm{p}}\right)\phi\right] = 0, \tag{5}$$

where $\delta$ is the Kronecker delta, which is also valid approximately in the weakly guiding regime with negligible SOC. (See Supplementary Note 3 for more detail) Equation (5) implies that $J_{\mathrm{T,OVM}}$ is always zero in the absence of SOC, as the SAM cannot be preserved ($\sigma_{\mathrm{s}} \neq 3\sigma_{\mathrm{p}}$), which corroborates that the SOC-induced SOE is crucial in facilitating the TH-OVG process.

The SOC-induced SOE is a confinement effect that generally strengthens as light is trapped in a smaller cross-section. To figure out the influence of the effect on TH-OVG more quantitatively, we calculate $J_{\mathrm{T,OVM}}$ for our experimental scenario where a 17-μm-thick cladding-etched silica optical fiber is tapered, as shown in Fig. 2(a). $J_{\mathrm{T,OVM}}$ increases monotonically as the cladding diameter falls. Such an increase of $J_{\mathrm{T,OVM}}$ can be contributed by two factors. One is the reduction of the effective mode areas [49] of the pump and third harmonic. This contribution also commonly affects other nonlinear optical effects, such as the self-phase modulation (SPM) of the pump and the pump-driven cross-phase modulation (XPM) of the third harmonic, monotonically increasing the respective overlap integrals $J_{\mathrm{P,OVM}}$ and $J_{\mathrm{X,OVM}}$. (See Supplementary Note 3 for the definitions of $J_{\mathrm{P,OVM}}$ and $J_{\mathrm{X,OVM}}$) However, the enhancement of $J_{\mathrm{T,OVM}}$ by a factor of ~6,000 during the entire course of tapering is highly substantial, compared to only ~100 or smaller for those of $J_{\mathrm{P,OVM}}$ and $J_{\mathrm{X,OVM}}$ and even that of $J_{\mathrm{T1}}$ for THG in the $HE_{12}$ mode [28,48]. (See Supplementary Note 2 for the definition of $J_{\mathrm{T1}}$) The dramatic enhancement of $J_{\mathrm{T,OVM}}$ arises dominantly from another contribution of the SOC-induced SOE, whereas the increases in other *J*'s are governed predominantly by the reduction in the effective mode area. We also calculate the canonical OAMs and SAMs of the pump and third-harmonic OVMs according to the cladding diameter. As the OVMs exhibit SOC and SOC-induced SOE more strongly, they become more deviated from the OAM/SAM eigenmodes for the paraxial regime, and their canonical OAMs and SAMs differ more from the nominal integer values [25,26]. Accordingly, we may employ the resulting change of the canonical OAM/SAM of the OVM as a quantitative measure of the SOC strength or the degree of SOE. As the fiber gets tapered down, the SOC-induced shift in the canonical OAM/SAM value increases monotonically, as shown in Fig. 2(b), which indicates the reinforcement of SOC. The strengthened SOC-induced SOE boosts up $J_{\mathrm{T,OVM}}$ cooperatively with the decrease in the effective mode area. We note that even in the presence of SOC each OVM is kept as a TAM eigenmode with a fixed integer TAM eigenvalue regardless of the cladding diameter variation. (See Supplementary Figure S4)

## Experimental results

### Observation and characterization of TH-OVG

For experimental observation of TH-OVG, we first fabricate silica multimode adiabatic submicron tapers (MASTs) from conventional telecom single-mode fiber (SMF) employing the recently developed two-step process [28]. We deeply wet-etch the SMF cladding below 20 μm and then taper the etched SMF down to submicron diameter using the conventional flame-brushing and pulling technique [50]. The MAST waist acts as a multimode nanofiber guiding spin-orbit-entangled OVMs, and we adjust the MAST waist diameter close to ~717 nm to achieve the phase matching for intermodal THG in the $\mathrm{OV}_{\pm 2,1}^{\pm}$ ($HE_{31}$) mode (Fig. 1(c)). The waist has a 5 mm length and is connected to untapered fiber sections via a pair of ~30-mm-long exponential transitions. During the whole tapering process, we check that the 532-nm $LP_{21}$ mode (a combination of the $HE_{31}$ and the $EH_{11}$ mode) transmits adiabatically through the fiber taper. (See Ref. [28] for the detailed description of MAST fabrication and its working principle) The final transmissions of the 532-nm $LP_{21}$ mode and the 1550-nm $LP_{01}$ mode are measured to be 58% and 94%, respectively. We package the MAST in an acrylic box to prevent its contamination for subsequent long-term experiments.

In our experiments, we use an all-fiber master oscillator power amplifier (MOPA) system as a pump source for THG (Fig. 3(a)). It incorporates a homemade wavelength-tunable narrow-linewidth passively mode-locked fiber laser that emits a 110 ps pulse train at a 2.1 MHz repetition rate over the wavelength tuning range of 1535–1563 nm [51]. We control the polarization state of the MOPA output with a fiber polarization controller and launch the pump beam into the fabricated MAST. We observe that the spatial profile of the third-harmonic output signal is generally in a mixture of multiple higher-order modes such as the $LP_{02}$, the $LP_{21}$, and a donut mode, depending on the polarization state of the pump beam. By adjusting the pump polarization state, we can make the third-harmonic output signal closely resemble the purely donut-shaped mode. Such the donut mode can be created with the highest purity at two mutually orthogonal pump polarization states. We measure the output power of the donut-shaped third-harmonic signal over the entire wavelength tuning range of the MOPA system (1535–1563 nm) at a fixed average input pump power of 0.25 W. The THG in the donut-shaped mode is most efficient at a specific pump wavelength (1543 nm for Fig. 3(b)), which is close to the design wavelength of 1550 nm, while other pump wavelengths also generate smaller amounts of donut-shaped third-harmonic signal (Fig. 3(c)) mainly due to the waist diameter nonuniformity [28].

We also measure the third-harmonic output powers over a range of pump powers, while fixing the pump wavelength at 1543 nm. The third-harmonic powers are almost identical for the two mutually orthogonal pump polarization states that yield the purest third-harmonic donut mode, exhibiting a theoretically predicted cubic dependence on the pump power, as shown in Fig. 3(d). The maximum THG conversion efficiencies are measured as $1.54\times10^{-6}$ and $1.58\times10^{-6}$ for the two pump polarizations at 0.37 W pump power. We measure the topological charge of the third-harmonic signal based on the astigmatic field transform through the beam focusing at a cylindrical lens (focal length: 50 mm) [52]. For one pump polarization state (denoted by 'A'), the transformed third-harmonic field displays a counter-clockwise slanted array of three lobes, as shown in Fig. 3(e), verifying together with the donut-shaped intensity profile that the generated third-harmonic field is an optical vortex with a topological charge of $l_s = +2$. On the contrary, for another pump polarization state (denoted by 'B'), an array of three lobes appears again but slants clockwise, which indicates an optical vortex with a topological charge of $l_s = -2$. These optical vortex structures are maintained over a broad range of pump powers, though their purities become slightly degraded as the pump power increases beyond ~0.37 W, mainly due to the unwanted THG in the $HE_{12}$ mode associated with the nonlinear pump polarization change [28,53].

The measured third-harmonic spectra in Figs. 3(f) and 3(g) show that the two optical vortices with the respective topological charges of +2 and -2 have identical wavelengths (514.4 nm), which almost equal one-third of the pump wavelength (1543 nm in Fig. 3(h)). As the pump power rises beyond ~0.37 W, a pair of sidebands appear in the spectra of both the pump and third harmonic. The sideband locations are the same for the third harmonic and the pump, distant from the main spectral peaks by ~2 THz. In addition, both the pump in the fundamental $\mathrm{OV}^{\pm}_{0,1}$ ($HE_{11}$) mode and the third harmonic in the $\mathrm{OV}^{\pm}_{\pm 2,1}$ ($HE_{31}$) mode are in the normal dispersion regime in the MAST waist. Therefore, the creation of the spectral sidebands can be attributed to XPM-induced modulation instability [28,54,55].

In principle, the third-harmonic OVMs of left-handed ($\mathrm{OV}^{+}_{+2,1}$) and right-handed ($\mathrm{OV}^{-}_{-2,1}$) quasi-circular polarizations are generated from the pump beam of left-handed ($\mathrm{OV}^{+}_{0,1}$) and right-handed ($\mathrm{OV}^{-}_{0,1}$) quasi-circular polarizations, respectively, in the MAST waist. In practice, however, the polarizations vary in the lead fiber sections to the MAST output and input ports due to the residual birefringence. Moreover, the residual stress and/or elliptical deformation can induce significant coupling of the $HE_{31}$ mode to the nearly degenerate $EH_{11}$ mode in the weakly guiding lead fiber [56]. These two effects hinder the precise experimental identification of the polarization states of the pump and third harmonic in the MAST waist. Nevertheless, we experimentally examine their polarization profiles at the output port using wave plates and a linear polarizer in addition to the existing setup in Fig. 3(a). We confirm that the polarization states of the two output third-harmonic optical vortices are always almost orthogonal to each other, and so are the corresponding pump polarization states simultaneously, although they generally appear elliptical. In addition, the polarization state of the output third harmonic is almost uniform over the entire beam cross-section, which supports that the third harmonic is close to an OVM made up of a balanced (50:50) superposition of the even and odd hybrid modes as in Eq. (1), rather than a vector beam with a nonuniform polarization distribution [57,58] consisting of a significantly unbalanced combination of the two hybrid modes. Finally, the topological charge of the third-harmonic output is measured as purely either +2 or -2 regardless of the elliptical polarization, which supports the existence of the mode mixing between the nearly degenerate $HE_{31}$ and $EH_{11}$ modes, whereas the circular birefringence is negligible. It is because such mode mixing always appears as a change of polarization (SAM) only, while preserving the OAM unless there is circular dichroism [56]. One might think of an alternative type of birefringence-induced mode coupling that occurs between the even and odd modes. However, such even-odd mode mixing gives rise to the simultaneous sign reversal of the OAM and SAM according to Eq. (1) and consequently yields a mixture of two third-harmonic OAMs of +2 and -2 at the output, which is contradictory to our experimental observation.

**Effect of modal birefringence in MAST on TH-OVG**

When the MAST cross-section is perfectly circular, the pump $OV_{0,1}^{+}$ and $OV_{0,1}^{-}$ modes that are left-handed and right-handed quasi-circularly polarized yield TH-OVG in the $OV_{+2,1}^{+}$ and $OV_{-2,1}^{-}$ mode, respectively, in a symmetric fashion while completely forbidding THG in the $OV_{0,2}^{\pm}$ ($HE_{12}$) mode. However, we observe in some cases that the third harmonic in the $HE_{12}$ mode is not entirely suppressed through the pump polarization control, which degrades the vortex purity and the symmetry of TH-OVG with respect to the handedness of pump polarization. We attribute the non-ideal TH-OVG performance to the non-circular MAST cross-section, which arises primarily in the deep wet-etching step of the MAST fabrication process. (See Supplementary Note 4 and Supplementary Figure 5 for more detail) The overall TH-OVG process can be viewed as a linear superposition of four THG processes involving different combinations of even/odd hybrid modes of the pump and third harmonic (2 for pump and 2 for third harmonic), as illustrated in Fig. 4(a). (See Supplementary Note 4 for more detail.) The TH-OVG then works properly when the even and odd hybrid modes are degenerate for both the pump and third harmonic. However, the deviation of the MAST cross-section from the circular one breaks the modal degeneracy, which hinders the perfect operation of TH-OVG.

We model the non-circular MAST cross-section as an elliptically deformed one, as illustrated in Fig. 4(b), and investigate the influence of the deformation-induced birefringence on TH-OVG based on the full-vectorial finite element analysis of the spatial modes in the MAST. First, we calculate the beat length between the even and odd hybrid modes for the pump and the third harmonic according to the eccentricity of the MAST cross-section. As the eccentricity rises, the degeneracy becomes broken more strongly, which appears as the decrease in the beat length, as shown in Figs. 4(c) and 4(d). In particular, the degeneracy is broken more seriously for the pump $HE_{11}$ mode and the third-harmonic $HE_{12}$ mode because their transverse electric fields are almost parallel to the major or minor axis of the cross-section [56]. The beat lengths can be even shorter than the MAST waist length (5 mm in our experiments) at relatively small eccentricities below 0.1. Then, the relative phase between the even and odd modes varies as the pump and third-harmonic fields propagate along the MAST, which can deteriorate the performance and controllability of TH-OVG. For instance, the THG in the $HE_{12}$ mode cannot be fully suppressed as the pump polarization state varies during propagation. Furthermore, while the four constituent THG processes in Fig. 4(a) are phase matched simultaneously for a perfectly circular MAST with degenerate even and odd hybrid modes, their phase-matching conditions become deviated from each other as the eccentricity rises (Figs. 4(e) and 4(f)). The phase-matching nanofiber diameters for the four processes of THG in the $HE_{12}$ mode can differ by even tens of nanometers. Such a large deviation of phase matching conditions also hinders the suppression of the THG in the $HE_{12}$ mode. In contrast, the third-harmonic $HE_{31}$ mode has a nonuniform polarization profile, and consequently, the degree of lifting the degeneracy is less significant, resulting in relatively longer beat lengths. Hence, the symmetry in the TH-OVG with respect to the pump polarization is more robust against the elliptical deformation compared to the vortex purity with the suppression of the THG in the $HE_{12}$ mode.

## Discussion

We have revealed that SOC-induced photonic SOE in a multimode optical nanofiber facilitates the efficient nonlinear frequency conversion from the spin-polarized Gaussian-like pump beam to the harmonic optical vortex with a nonzero OAM. This remarkable observation contrasts sharply with a myriad of previous experiments employing conventional isotropic media, where TH-OVG is forbidden. Furthermore, our full-vectorial nonlinear coupled-mode theory unveils the crucial role of SOE in creating spin-to-orbit-converting nonlinear polarization, which offers a unique and powerful means of all-optical generation of an optical vortex without the requirement of any structured pump beams, sophisticatedly designed media or particular types of materials. We emphasize that our MAST is a unique and ideal platform for spin-orbit-interacting nonlinear optics, as providing adiabatic guidance of multiple spatial modes, including spin-orbit-entangled OVMs, together with the ultrahigh nonlinearity and broad dispersion controllability owing to the tight light confinement. All these features are essential to achieve phase-matched efficient nonlinear frequency conversion among different OVMs.

The new research paradigm of spin-orbit-interacting nonlinear wave mixing of spin-orbit-entangled light is not limited to TH-OVG. It can also be extended to other intermodal nonlinear frequency conversion processes, such as different kinds of harmonic generation [59–61], FWM [62], and stimulated Raman scattering [63], as well as a combination of two or more nonlinear optical processes, which could generate optical vortices of topological charges other than ±2 or in the superposition/entanglement of different topological charges. We also anticipate that a coherent acoustic vortex might be generated in an optical nanofiber via stimulated Brillouin scattering by pumping with a spin-orbit-entangled OVM [27]. Conversely, exciting and controlling torsional acoustic vibration in an optical nanofiber [64,65] could implement nonlinear switching of chirality and optical vortices. Furthermore, recently realized exotic kinds of structured light, such as polarization vortices [57,58], polarization Möbius strips [66], optical vortex knots [67], non-integer vortices [68], optical vortices with multiple phase singularities [69], and optical wheels having transverse OAMs [70], might also be all-optically created and manipulated via spin-orbit-coupling nonlinear wave mixing.

Although residual THG in the unwanted $HE_{12}$ mode currently degrades the vortex purity, we believe that such mixing of undesired modes might be resolved by employing specially designed optical waveguides with proper modal dispersions, such as vortex fibers [64], microstructured fibers [71], and integrated on-chip waveguides [72]. In addition, the TH-OVG conversion efficiency achieved so far is on the order of $\sim 10^{-6}$, limited primarily by short effective interaction length. The effective indices of tightly confined modes are sensitive to the waveguide geometry, the nanofiber diameter variation as small as only a few nanometers disturbing the phase matching significantly [28,73]. The nanofibers made out of other glasses with larger optical nonlinearities (e.g., chalcogenide glass [74]) or the existing silica MAST coated with such materials [75] might enhance the conversion efficiency.

Finally, it is worthwhile to mention some potential key applications of our work. As nonlinear optical frequency conversion processes usually produce strong correlation among multiple photonic degrees of freedom (e.g., frequency, propagation direction, position/timing, as well as SAM/OAM) of different optical frequencies owing to the requirement of conservation of physical quantities (i.e., total energy and wavevector, as well as TAM). Hence, our scheme could immediately apply to multi-dimensional photonic quantum information processing. In addition, the guided light along the MAST waist can efficiently interact with other photonic systems in the vicinity via the evanescent fields. The visible optical vortices generated via TH-OVG in the MAST waist, together with the quasi-circularly polarized pump beam at the telecom wavelengths, provide powerful tools for sophisticated manipulation of atoms/molecules or nanoparticles trapped near the MAST waist via their strong evanescent fields [76,77]. Moreover, the harmonic optical vortex can be coupled to various nanophotonic systems [78–81] for efficient photonic information processing. In recent noticeable experiments [82], for instance, the guided light is coupled to long-lived quantum memories (e.g., nitrogen-vacancy centers) imbedded in optical nanofibers for storage and retrieval of the information encoded in the polarization state of telecom photons via THG. Our work might be combined with conventional telecom optical networks [83] to implement multi-dimensional hybrid quantum networks, where the OAMs of visible photons are connected to the SAMs of telecom photons and the quantum memories.

## Acknowledgements

This work was supported by the National Research Foundation of Korea (NRF) grants funded by the Korea government (NRF-2020K1A3A1A19088178, NRF-2022R1H1A2092792) and the KAIST $C^2$ Project.

## Competing interests

The authors declare no competing interests.

## Materials & correspondence

Correspondence and requests for materials should be addressed to M.S.K.

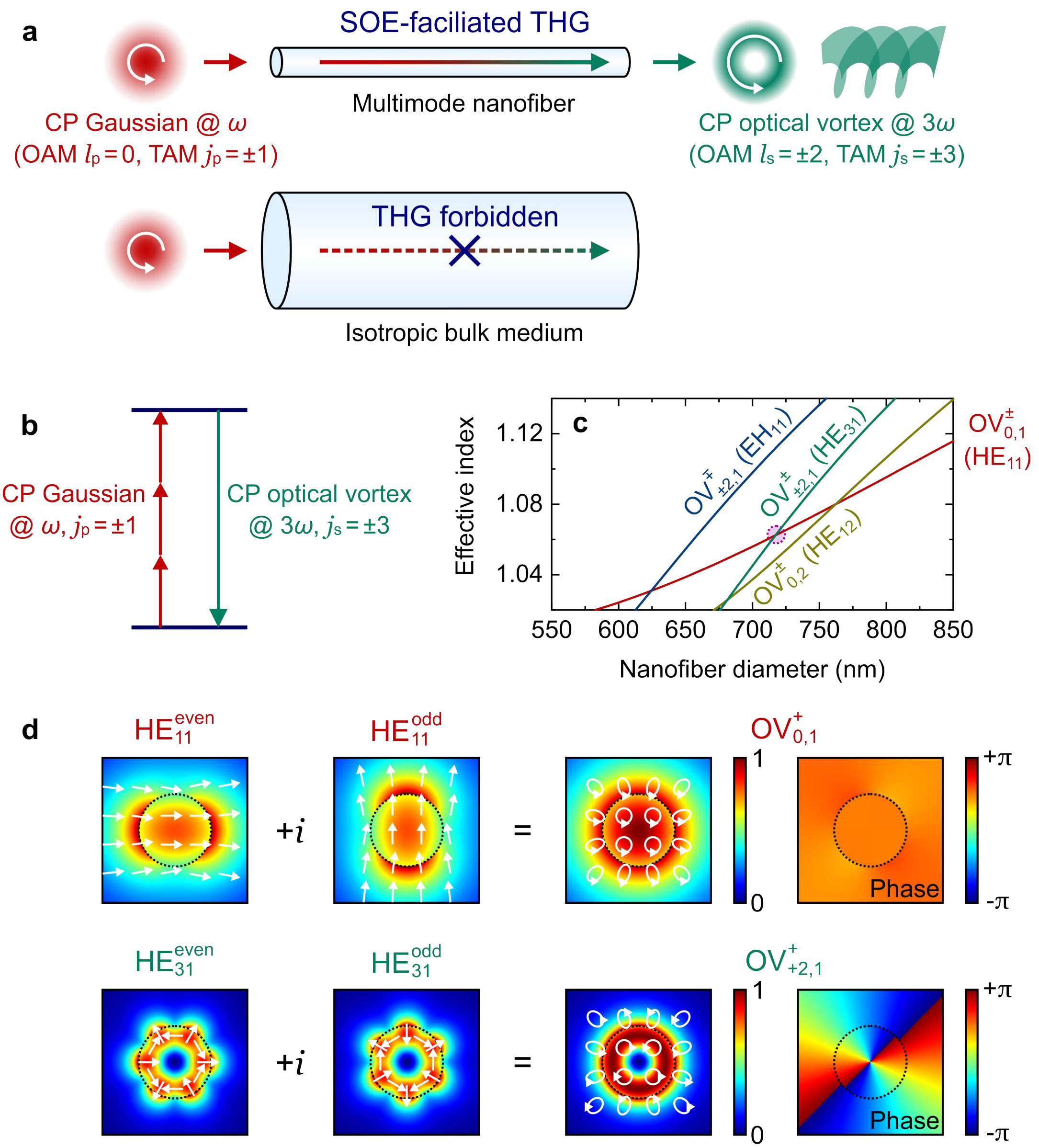

**FIG. 1. Concept and principle of third-harmonic optical vortex generation (TH-OVG) facilitated by photonic spin-orbit entanglement (SOE).** (**a**) Schematic diagram of the proposed TH-OVG process. The circularly polarized (CP) Gaussian-like pump beam at optical frequency $\omega$ is converted into a CP optical vortex at optical frequency $3\omega$ via third-harmonic generation (THG) in a multimode nanofiber. The input pump beam has no orbital angular momentum (OAM, $l_p = 0$) but only a spin of $\sigma_p = \pm1$, so its total angular momentum (TAM) is $j_p = \pm1$. The output third harmonic has an OAM of $l_s = \pm2$ and a spin of ±1 with the same handedness, so its TAM is $j_s = \pm3$. This TH-OVG process is facilitated by the SOE manifested in the pump and the third harmonic, which arises from the confinement-induced spin-orbit coupling in the multimode nanofiber. In sharp contrast, the TH-OVG is forbidden in isotropic bulk media, as THG cannot occur with a CP pump beam. (**b**) Energy level diagram for TH-

OVG displaying the simultaneous conservation of energy and TAM in the process. (**c**) Calculated effective indices of some optical vortex modes (OVMs) in silica-glass multimode nanofibers as a function of the nanofiber diameter. The red curve corresponds to the pump beam in the fundamental OVM ( $\mathrm{OV}_{0,1}^{\pm} = \mathrm{HE}_{11}^{\mathrm{even}} \pm i\,\mathrm{HE}_{11}^{\mathrm{odd}}$ ) at 1550 nm wavelength, whereas the others are the third harmonic in the higher-order OVMs. (**d**) Pump and third-harmonic OVMs that are phase-matched for the TH-OVG in the $\mathrm{OV}_{\pm 2,1}^{\pm}$ mode in a 717-nm-thick nanofiber indicated by the magenta dashed circle in (**c**). The profiles of the field amplitude and phase of each OVM are shown, together with the field profiles of the constituent even and odd hybrid modes—the pump $HE_{11}$ modes and the third-harmonic $HE_{31}$ modes. The white arrows and black dashed circles represent the polarization profile and the nanofiber cross-section, respectively. The phase pattern in the rightmost column displays the phase of the horizontal (x) field component relative to the longitudinal (z) one. See Supplementary Note 1 for the case of right-handed CP OVMs.

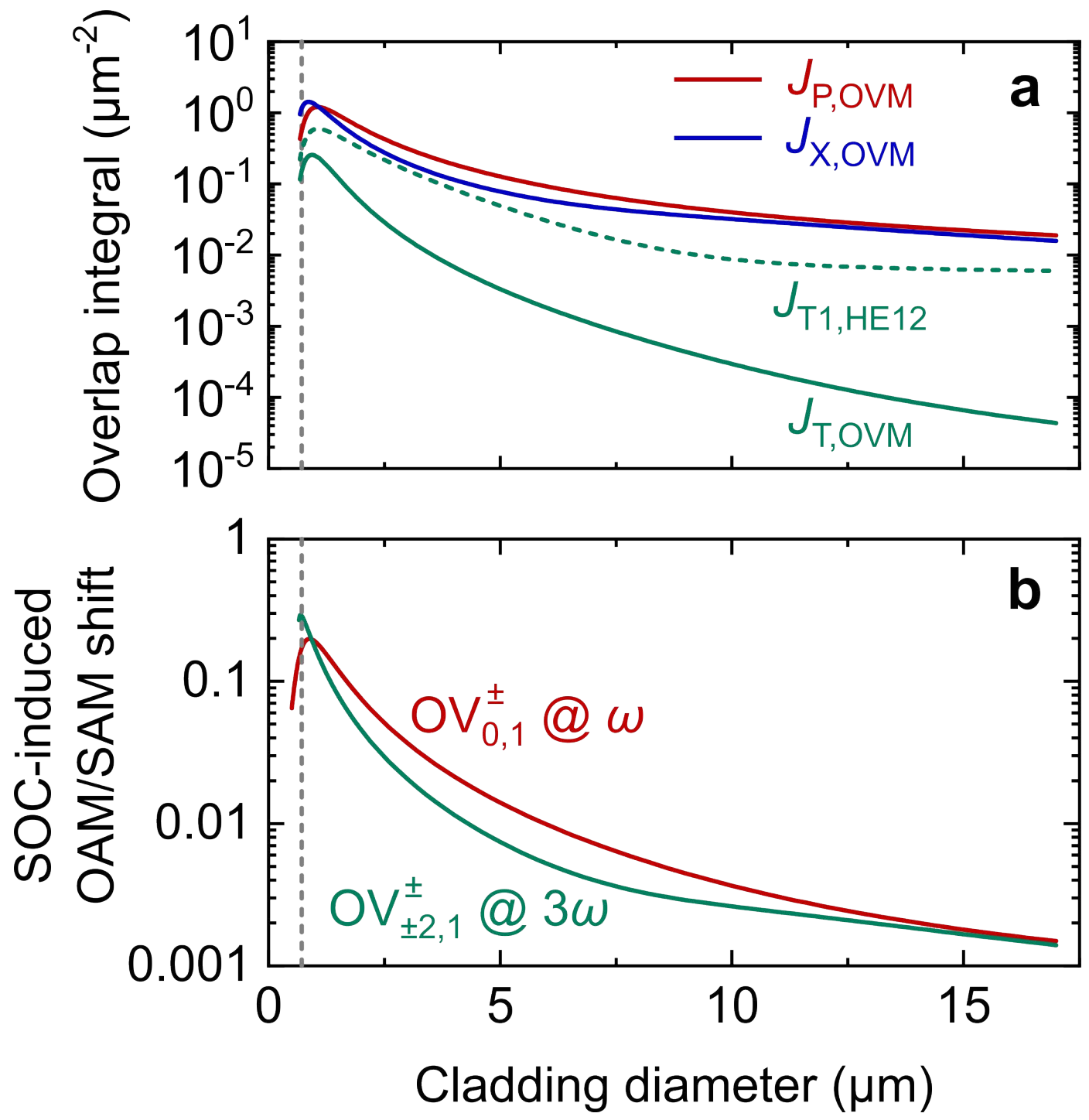

**FIG. 2. Theoretical analysis of spin-orbit coupling (SOC) and third-harmonic optical vortex generation (TH-OVG) in optical microfibers and nanofibers.** (**a**) Calculated overlap integrals for third-order nonlinear optical processes involving optical vortex modes (OVMs) according to the cladding diameter in the course of tapering of an etched silica optical fiber. The turquoise solid curve ($J_{\mathrm{T,OVM}}$) corresponds to TH-OVG in the $\mathrm{OV}^{\pm}_{\pm2,1}$ OVM from the 1550 nm pump in the $\mathrm{OV}^{\pm}_{0,1}$ OVM. The red curve ($J_{\mathrm{P,OVM}}$) corresponds to the self-phase modulation of the pump OVM, and the blue one ($J_{\mathrm{X,OVM}}$) to the cross-phase modulation of the third-harmonic OVM by the pump OVM. The three overlap integrals are defined in Eqs. (S28) and (S29) in Supplementary Note 3. On the other hand, the turquoise dashed curve ($J_{\mathrm{T1,HE12}}$ defined in Eq. (S14) in Supplementary Note 2) represents the third-harmonic generation (THG) in the $\mathrm{HE}_{12}$ hybrid mode from the pump in the $\mathrm{HE}_{11}$ hybrid mode of the same even/odd parity [28]. Note that $J_{\mathrm{T,OVM}} = 0$ for the THG in the $\mathrm{OV}^{\pm}_{0,2} = \mathrm{HE}^{\mathrm{even}}_{12} \pm i\,\mathrm{HE}^{\mathrm{odd}}_{12}$ mode, so it does not appear in this plot of the current logarithmic vertical scale. (**b**) Calculated SOC-induced change of the canonical orbital angular momenta (OAMs) of the $\mathrm{OV}^{\pm}_{0,1}$ mode at the 1550 nm pump wavelength and the $\mathrm{OV}^{\pm}_{\pm2,1}$ mode at the third harmonic, according to the cladding diameter. The SOC-induced changes of the canonical spin angular momenta (SAMs) occur with the opposite signs, so the total angular momenta are preserved regardless of the cladding diameter variation. See Supplementary Note 3 and Supplementary Figure 4 for more detail. In both (**a**)

and (**b**), the initial core [cladding] diameter and numerical aperture of the etched fiber prior to tapering are set as the experimental values of 8.7 [17] μm and 0.13, respectively, and the grey vertical dashed lines indicate the phase-matching nanofiber diameter of 717 nm for the TH-OVG process.

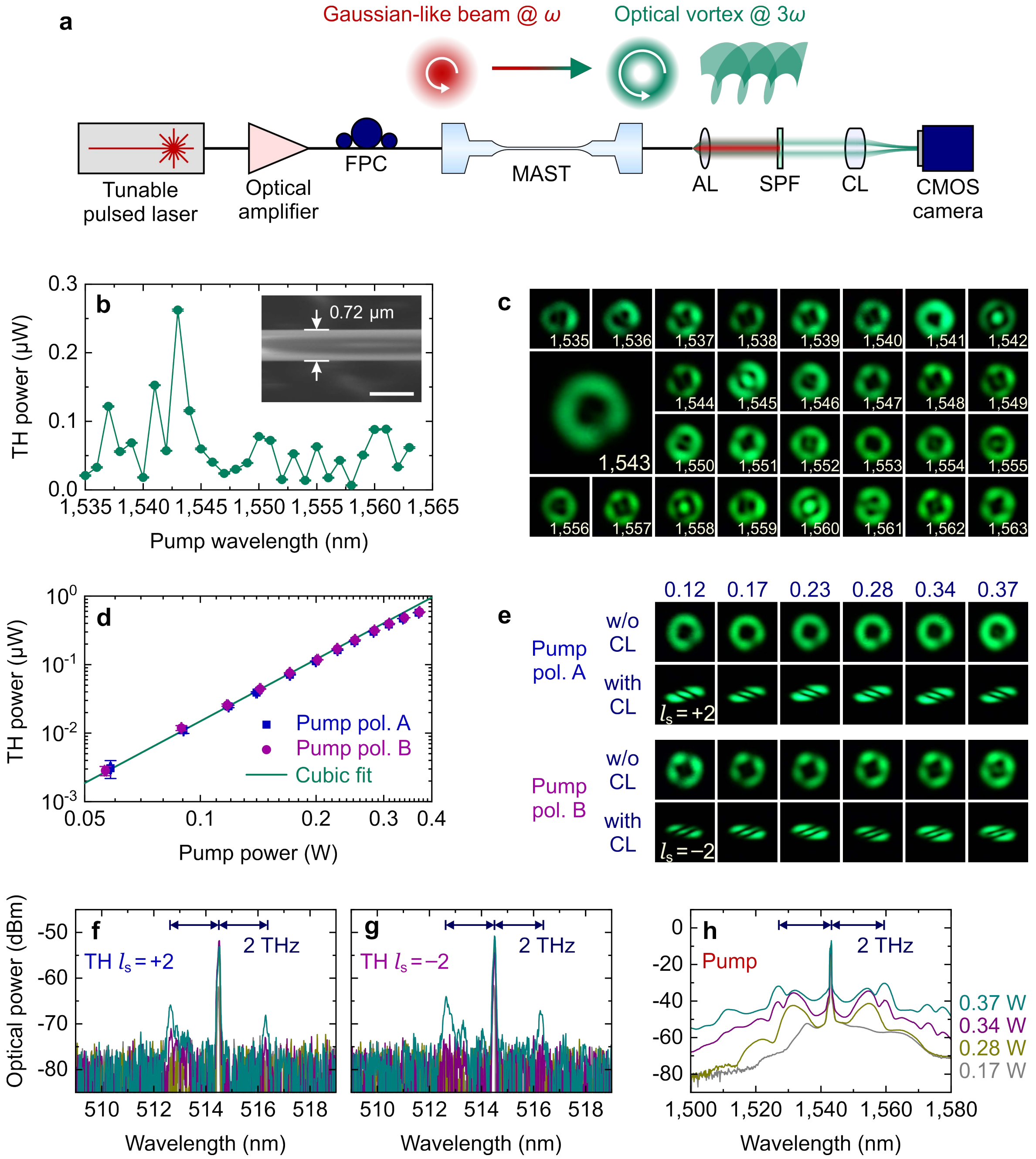

**FIG. 3. Experimental observation of third-harmonic optical vortex generation (TH-OVG).** (**a**) Schematic diagram of the experimental setup. The output from a homemade wavelength-tunable mode-locked fiber laser [51] is amplified by an erbium-doped-fiber optical amplifier. This pump beam is launched into a multimode adiabatic submicron taper (MAST) after its polarization is adjusted with a fiber polarization controller (FPC) to generate a third-harmonic optical vortex. AL, aspheric lens; SPF, short-pass filter; CL, cylindrical lens; CMOS, complementary metal-oxide-semiconductor. (**b**) Measured third-harmonic output powers at a fixed average pump power of 0.25 mW over the pump wavelength tuning range of 1535–1563

nm. The polarization state at each pump wavelength is adjusted to minimize the THG in the $HE_{12}$ mode. The inset shows the scanning electron micrograph of the MAST waist, where the white scale bar corresponds to 1 μm. (**c**) Far-field profiles of the third-harmonic output observed without the CL at different pump wavelengths (white numbers in nanometers). (**d**) Measured third-harmonic output powers over a range of average pump powers at a fixed pump wavelength of 1543 nm yielding the maximum TH-OVG conversion efficiency, as seen in (**b**). The blue and purple correspond to the two mutually orthogonal pump polarization states that yield optimum TH-OVG, which we denote by pump polarization A and B. The turquoise line is the cubic fit to the measurements. (**e**) Far-filed profiles of the third-harmonic output observed with the pump polarization states A and B without and with the CL. The white numbers indicate the average pump powers in watts. The far-field profiles obtained with the CL reveal the topological charges $l_s$ = +2 (upper) and -2 (lower) of the third-harmonic output signals. (**f**,**g**) Optical spectra of the third-harmonic output signals with $l_s$ = +2 (**f**) and -2 (**g**) obtained at pump polarization states A and B, respectively. (**h**) Optical spectra of the output pump beam. In (**f**–**h**), the color of each curve corresponds to the average pump power displayed in (**h**), 0.17 (grey), 0.28 (dark yellow), 0.34 (dark magenta), and 0.37 W (dark cyan).

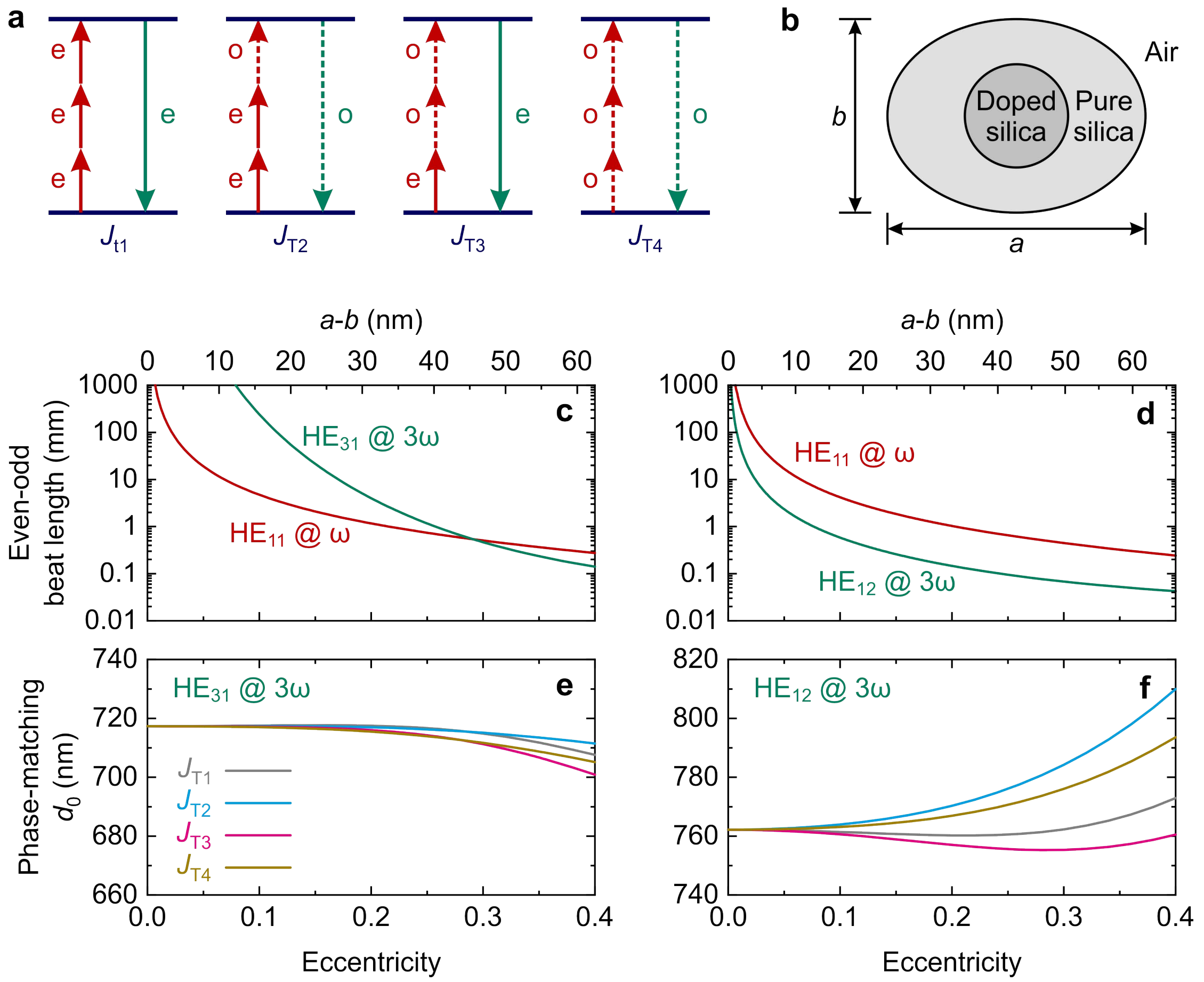

**FIG. 4. Effect of the deformation-induced nanofiber birefringence on the phase matching for third-harmonic optical vortex generation.** (**a**) Energy level diagrams representing four third-harmonic generation (THG) processes that involve different combinations of the pump and third harmonic hybrid modes, where $J_{Ti}$'s ($i$ = 1 to 4) are the corresponding overlap integrals. (See Supplementary Note 2 for their definitions) The red and turquoise stand for the pump and third harmonic, and the solid ('e') and dashed ('o') arrows represent the even and odd hybrid modes. (**b**) Model of the cross-section of birefringent nanofiber with elliptical deformation. The wet-etched pure-silica cladding is elliptical, whereas the germanium-doped-silica core is perfectly circular. $a$ and $b$ are the major and minor axis, respectively, of the elliptical cladding. (**c**,**d**) Calculated beat lengths between the even and odd hybrid modes for the pump $HE_{11}$ mode (red curves) and the third harmonic $HE_{31}$ (**c**) and $HE_{12}$ (**d**) modes (turquoise curves) over a range of cladding eccentricity defined as $\sqrt{1-(b/a)^2}$ . The arithmetic mean of $a$ and $b$, $d_0 = (a+b)/2$, is fixed as the phase-matching diameters of 717 and 762 nm for THG in the $HE_{31}$ and the $HE_{12}$ mode, respectively, in the perfectly circular nanofibers. (**e**,**f**) Calculated phase-matching $d_0$ for THG in the $HE_{31}$ (**e**) and the $HE_{12}$ mode (**f**) with four different

combinations of even/odd hybrid modes defined in (**a**) over a range of cladding eccentricity. In (**c–f**), the pump wavelength is fixed at 1550 nm, and the diameter of the germanium-doped-silica core is kept smaller than $d_0$ by a factor of 8.7/17, the same as the experimental value.

# Supplementary Information

## Supplementary Note 1: Optical vortex modes (OVMs) in optical nanofiber

An optical waveguide with a circular cross-section supports OVMs as the guided eigenmodes. Each OVM designated by $\mathrm{OV}_{l,m}^{\sigma}$ can be formed by linear superposition of the even and odd hybrid modes with the phase difference of 90º as below [1,2].

$$\begin{aligned}\mathrm{OV}_{\pm l,m}^{\pm} &= \mathrm{HE}_{l+1,m}^{\mathrm{even}} \pm i\,\mathrm{HE}_{l+1,m}^{\mathrm{odd}},\\ \mathrm{OV}_{\pm l,m}^{\mp} &= \mathrm{EH}_{l-1,m}^{\mathrm{even}} \pm i\,\mathrm{EH}_{l-1,m}^{\mathrm{odd}},\end{aligned} \tag{S1}$$

where $l$ and $\sigma$ represent the canonical orbital angular momentum (OAM) and spin angular momentum (SAM) numbers, respectively, under the paraxial approximation, and $m$ is the radial mode number. Figure S1 displays the electric field profiles of four OVMs, the $\mathrm{OV}_{0,1}^{\pm} = \mathrm{HE}_{1,1}^{\mathrm{even}} \pm i\,\mathrm{HE}_{1,1}^{\mathrm{odd}}$ modes at 1550 nm wavelength and the $\mathrm{OV}_{2,1}^{\pm} = \mathrm{HE}_{3,1}^{\mathrm{even}} \pm i\,\mathrm{HE}_{3,1}^{\mathrm{odd}}$ modes at the third harmonic, in a silica glass optical nanofiber of 717 nm diameter, where the OVMs are phase matched for third-harmonic optical vortex generation. The $\mathrm{OV}_{\pm l,m}^{\pm}$ modes composed of the HE hybrid modes have the OAM and SAM in the same handedness, so they are referred to as the spin-orbit aligned modes in some literature [3]. On the contrary, the $\mathrm{OV}_{\pm l,m}^{\mp}$ modes consisting of the EH hybrid modes are the spin-orbit anti-aligned modes, in the sense that the OAM and SAM are in the opposite handedness.

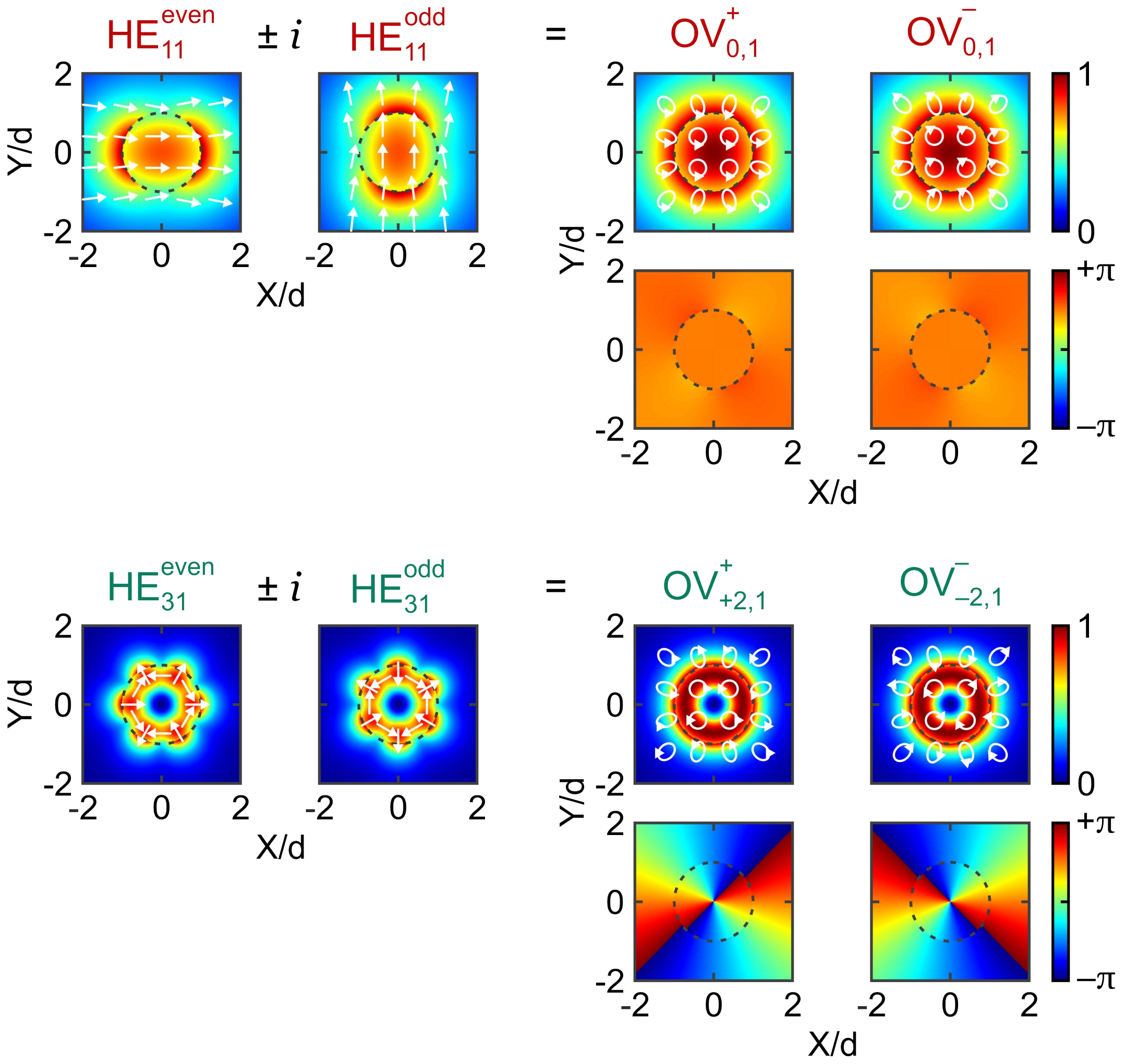

**FIG. S1. Electric field profiles of four optical vortex modes (OVMs) in an optical nanofiber.** The $\mathrm{OV}_{0,1}^{\pm} = \mathrm{HE}_{1,1}^{\mathrm{even}} \pm i\,\mathrm{HE}_{1,1}^{\mathrm{odd}}$ modes at 1550 nm wavelength and the $\mathrm{OV}_{2,1}^{\pm} = \mathrm{HE}_{3,1}^{\mathrm{even}} \pm i\,\mathrm{HE}_{3,1}^{\mathrm{odd}}$ modes at the third harmonic in a silica glass optical nanofiber with the diameter of 2$d$ = 717 nm are displayed. The dark grey dashed circles correspond to the air-silica interface. The white arrows indicate the polarization distributions and are superimposed upon color maps of normalized field amplitudes. The color maps without any white arrows show the profiles of the phase of the x-field component relative to the z-field component on the +x-axis.

**Supplementary Note 2: Derivation of the full-vectorial nonlinear coupled-mode equations describing the intermodally phase-matched third-harmonic generation (THG)**

We derive the full-vectorial nonlinear coupled-mode equations that describe the experimentally observed intermodally phase-matched THG process. Here, we treat the pump field ($\mathbf{E}_\text{p}$ and $\mathbf{H}_\text{p}$) and the third-harmonic field ($\mathbf{E}_\text{s}$ and $\mathbf{H}_\text{s}$) propagating along the z-axis as linear combinations of the corresponding even and odd hybrid modes as below.

$$\begin{aligned}
\mathbf{E}_\text{p}(\mathbf{r};\omega) &= Z_0^{1/2}\left[\mathbf{e}_\text{p,e}(x,y)a_\text{p,e}(z)e^{i\beta_\text{p,e}z} + \mathbf{e}_\text{p,o}(x,y)a_\text{p,o}(z)e^{i\beta_\text{p,o}z}\right],\\
\mathbf{H}_\text{p}(\mathbf{r};\omega) &= Z_0^{-1/2}\left[\mathbf{h}_\text{p,e}(x,y)a_\text{p,e}(z)e^{i\beta_\text{p,e}z} + \mathbf{h}_\text{p,o}(x,y)a_\text{p,o}(z)e^{i\beta_\text{p,o}z}\right],\\
\mathbf{E}_\text{s}(\mathbf{r};3\omega) &= Z_0^{1/2}\left[\mathbf{e}_\text{s,e}(x,y)a_\text{s,e}(z)e^{i\beta_\text{s,e}z} + \mathbf{e}_\text{s,o}(x,y)a_\text{s,o}(z)e^{i\beta_\text{s,o}z}\right],\\
\mathbf{H}_\text{s}(\mathbf{r};3\omega) &= Z_0^{-1/2}\left[\mathbf{h}_\text{s,e}(x,y)a_\text{s,e}(z)e^{i\beta_\text{s,e}z} + \mathbf{h}_\text{s,o}(x,y)a_\text{s,o}(z)e^{i\beta_\text{s,o}z}\right],
\end{aligned} \tag{S2}$$

$$\begin{aligned}
\mathbf{E}(\mathbf{r},t) &= \frac{1}{2}\left(\mathbf{E}_\text{p}(\mathbf{r};\omega)e^{-i\omega t} + \mathbf{E}_\text{s}(\mathbf{r};3\omega)e^{-i3\omega t} + \text{c.c.}\right),\\
\mathbf{H}(\mathbf{r},t) &= \frac{1}{2}\left(\mathbf{H}_\text{p}(\mathbf{r};\omega)e^{-i\omega t} + \mathbf{H}_\text{s}(\mathbf{r};3\omega)e^{-i3\omega t} + \text{c.c.}\right),
\end{aligned} \tag{S3}$$

where $\mathbf{E}(\mathbf{r},t)$ and $\mathbf{H}(\mathbf{r},t)$ are the total electric and magnetic fields, $\omega$ is the optical angular frequency of the pump field, and $Z_0 = 1/(\varepsilon_0 c)$ is the impedance of free space, where $\varepsilon_0$ and $c$ are the electric permittivity and speed of light, respectively, in a vacuum. $\mathbf{e}_q(x,y)$ and $\mathbf{h}_q(x,y)$ are the normalized electric and magnetic field distributions of the mode $q$, and $a_q(z)$ and $\beta_q$ are the slowly varying complex field amplitude and propagation constant of the mode. The subscripts p and s represent the pump and third-harmonic fields, and e and o stand for the even and odd modes. By substituting Eqs. (S2) and (S3) into Maxwell's equations and applying the reciprocity theorem for electromagnetic fields [4], we obtain

$$
\begin{aligned}
\frac{da_{\mathrm{p,e}}(z)}{dz} &= iZ_0^{1/2}\frac{\omega}{4}e^{-i\beta_{\mathrm{p,e}}z}\int\left(\mathbf{e}_{\mathrm{p,e}}^{*}\cdot\mathbf{P}_{\mathrm{p}}^{(3)}(\mathbf{r};\omega)\right)dA,\\
\frac{da_{\mathrm{p,o}}(z)}{dz} &= iZ_0^{1/2}\frac{\omega}{4}e^{-i\beta_{\mathrm{p,o}}z}\int\left(\mathbf{e}_{\mathrm{p,o}}^{*}\cdot\mathbf{P}_{\mathrm{p}}^{(3)}(\mathbf{r};\omega)\right)dA,\\
\frac{da_{\mathrm{s,e}}(z)}{dz} &= iZ_0^{1/2}\frac{3\omega}{4}e^{-i\beta_{\mathrm{s,e}}z}\int\left(\mathbf{e}_{\mathrm{s,e}}^{*}\cdot\mathbf{P}_{\mathrm{s}}^{(3)}(\mathbf{r};3\omega)\right)dA,\\
\frac{da_{\mathrm{s,o}}(z)}{dz} &= iZ_0^{1/2}\frac{3\omega}{4}e^{-i\beta_{\mathrm{s,o}}z}\int\left(\mathbf{e}_{\mathrm{s,o}}^{*}\cdot\mathbf{P}_{\mathrm{s}}^{(3)}(\mathbf{r};3\omega)\right)dA,
\end{aligned}
\tag{S4}
$$

$$
\mathbf{P}_{\mathrm{NL}}^{(3)}(\mathbf{r},t) = \frac{1}{2}\left(\mathbf{P}_{\mathrm{p}}^{(3)}(\mathbf{r};\omega)e^{-i\omega t} + \mathbf{P}_{\mathrm{s}}^{(3)}(\mathbf{r};3\omega)e^{-i3\omega t} + \mathrm{c.c.}\right),
\tag{S5}
$$

where we neglect the dispersion terms as we are interested in the (quasi-)continuous-wave regime. The surface integrals in Eq. (S4) and all the following equations are performed within the glass part of the waveguide cross-section. $\mathbf{P}_{\mathrm{NL}}^{(3)}(\mathbf{r},t)$ is the third-order nonlinear polarization, which brings about various types of third-order nonlinear optical processes involving the four hybrid modes (2 for the pump and 2 for the third harmonic). We consider self-phase modulation (SPM), cross-phase modulation (XPM), and four-wave mixing (FWM), as well as THG and degenerate third-order parametric down-conversion (TPDC). We obtain $\mathbf{P}_{\mathrm{p,s}}^{(3)}$ for THG and degenerate TPDC, which take the form of

$$
\begin{aligned}
\mathbf{P}_{\mathrm{p,TPDC}}^{(3)}(\mathbf{r};\omega) &= \frac{3}{4}\varepsilon_0\boldsymbol{\chi}^{(3)}(\omega;3\omega,-\omega,-\omega)\mathbf{E}_{\mathrm{s}}(\mathbf{r})\mathbf{E}_{\mathrm{p}}^{*}(\mathbf{r})\mathbf{E}_{\mathrm{p}}^{*}(\mathbf{r})\\
&= \frac{1}{4}\varepsilon_0\chi_{xxxx}^{(3)}\left[2\left(\mathbf{E}_{\mathrm{s}}\cdot\mathbf{E}_{\mathrm{p}}^{*}\right)\mathbf{E}_{\mathrm{p}}^{*} + \left(\mathbf{E}_{\mathrm{p}}^{*}\cdot\mathbf{E}_{\mathrm{p}}^{*}\right)\mathbf{E}_{\mathrm{s}}\right], \quad \text{(pump)}\\
\mathbf{P}_{\mathrm{s,THG}}^{(3)}(\mathbf{r};3\omega) &= \frac{1}{4}\varepsilon_0\boldsymbol{\chi}^{(3)}(3\omega;\omega,\omega,\omega)\mathbf{E}_{\mathrm{p}}(\mathbf{r})\mathbf{E}_{\mathrm{p}}(\mathbf{r})\mathbf{E}_{\mathrm{p}}(\mathbf{r})\\
&= \frac{1}{4}\varepsilon_0\chi_{xxxx}^{(3)}\left(\mathbf{E}_{\mathrm{p}}\cdot\mathbf{E}_{\mathrm{p}}\right)\mathbf{E}_{\mathrm{p}}. \quad \text{(third harmonic)}
\end{aligned}
\tag{S6}
$$

$\mathbf{P}_{\mathrm{p,s}}^{(3)}$ for SPM is written as [5]

$$
\begin{aligned}
\mathbf{P}^{(3)}_{\mathrm{p,SPM}}(\mathbf{r};\omega) &= \frac{3}{4}\varepsilon_0\boldsymbol{\chi}^{(3)}(\omega;\omega,\omega,-\omega)\mathbf{E}_{\mathrm{p}}(\mathbf{r})\mathbf{E}_{\mathrm{p}}(\mathbf{r})\mathbf{E}^*_{\mathrm{p}}(\mathbf{r}) \\
&= \frac{1}{4}\varepsilon_0\chi^{(3)}_{xxxx}\left[2\left(\mathbf{E}_{\mathrm{p}}\cdot\mathbf{E}^*_{\mathrm{p}}\right)\mathbf{E}_{\mathrm{p}} + \left(\mathbf{E}_{\mathrm{p}}\cdot\mathbf{E}_{\mathrm{p}}\right)\mathbf{E}^*_{\mathrm{p}}\right], \quad \text{(pump)} \\
\mathbf{P}^{(3)}_{\mathrm{s,SPM}}(\mathbf{r};3\omega) &= \frac{3}{4}\varepsilon_0\boldsymbol{\chi}^{(3)}(3\omega;3\omega,3\omega,-3\omega)\mathbf{E}_{\mathrm{s}}(\mathbf{r})\mathbf{E}_{\mathrm{s}}(\mathbf{r})\mathbf{E}^*_{\mathrm{s}}(\mathbf{r}) \\
&= \frac{1}{4}\varepsilon_0\chi^{(3)}_{xxxx}\left[2\left(\mathbf{E}_{\mathrm{s}}\cdot\mathbf{E}^*_{\mathrm{s}}\right)\mathbf{E}_{\mathrm{s}} + \left(\mathbf{E}_{\mathrm{s}}\cdot\mathbf{E}_{\mathrm{s}}\right)\mathbf{E}^*_{\mathrm{s}}\right], \quad \text{(third harmonic)}
\end{aligned}
\tag{S7}
$$

whereas for XPM combined with FWM [5],

$$
\begin{aligned}
\mathbf{P}^{(3)}_{\mathrm{p,XPM}}(\mathbf{r};\omega) &= \frac{3}{2}\varepsilon_0\boldsymbol{\chi}^{(3)}(\omega;3\omega,-3\omega,\omega)\mathbf{E}_{\mathrm{s}}(\mathbf{r})\mathbf{E}^*_{\mathrm{s}}(\mathbf{r})\mathbf{E}_{\mathrm{p}}(\mathbf{r}) \\
&= \frac{1}{2}\varepsilon_0\chi^{(3)}_{xxxx}\left[\left(\mathbf{E}_{\mathrm{s}}\cdot\mathbf{E}^*_{\mathrm{s}}\right)\mathbf{E}_{\mathrm{p}} + \left(\mathbf{E}_{\mathrm{p}}\cdot\mathbf{E}^*_{\mathrm{s}}\right)\mathbf{E}_{\mathrm{s}} + \left(\mathbf{E}_{\mathrm{p}}\cdot\mathbf{E}_{\mathrm{s}}\right)\mathbf{E}^*_{\mathrm{s}}\right], \quad \text{(pump)} \\
\mathbf{P}^{(3)}_{\mathrm{s,XPM}}(\mathbf{r};3\omega) &= \frac{3}{2}\varepsilon_0\boldsymbol{\chi}^{(3)}(3\omega;\omega,-\omega,3\omega)\mathbf{E}_{\mathrm{p}}(\mathbf{r})\mathbf{E}^*_{\mathrm{p}}(\mathbf{r})\mathbf{E}_{\mathrm{s}}(\mathbf{r}) \\
&= \frac{1}{2}\varepsilon_0\chi^{(3)}_{xxxx}\left[\left(\mathbf{E}_{\mathrm{p}}\cdot\mathbf{E}^*_{\mathrm{p}}\right)\mathbf{E}_{\mathrm{s}} + \left(\mathbf{E}_{\mathrm{s}}\cdot\mathbf{E}^*_{\mathrm{p}}\right)\mathbf{E}_{\mathrm{p}} + \left(\mathbf{E}_{\mathrm{s}}\cdot\mathbf{E}_{\mathrm{p}}\right)\mathbf{E}^*_{\mathrm{p}}\right]. \quad \text{(third harmonic)}
\end{aligned}
\tag{S8}
$$

Here, $\boldsymbol{\chi}^{(3)}$ is the third-order nonlinear susceptibility tensor. In the derivation of Eqs. (S6)–(S8), we utilize the fact that for waveguides made of isotropic materials such as fused silica glass, the elements of $\boldsymbol{\chi}^{(3)}$ that originates from non-resonant electronic response have the following properties [5].

$$
\begin{gathered}
\chi^{(3)}_{ijkl} = \chi^{(3)}_{xxyy}\delta_{ij}\delta_{kl} + \chi^{(3)}_{xyxy}\delta_{ik}\delta_{jl} + \chi^{(3)}_{xyyx}\delta_{il}\delta_{jk} \qquad (i,j,k,l = x,y,z), \\
\chi^{(3)}_{xxyy} = \chi^{(3)}_{xyxy} = \chi^{(3)}_{xyyx} = \chi^{(3)}_{xxxx}\big/3,
\end{gathered}
\tag{S9}
$$

where $\delta$ is the Kronecker delta. Furthermore, in the absence of birefringence, we can set $\beta_{\mathrm{p,e}} = \beta_{\mathrm{p,o}} \equiv \beta_{\mathrm{p}}$ and $\beta_{\mathrm{s,e}} = \beta_{\mathrm{s,o}} \equiv \beta_{\mathrm{s}}$. By inserting Eqs. (S6)–(S8) into Eq. (S4), we obtain the full-vectorial nonlinear coupled-mode equations for the slowly varying complex field amplitudes as follows.

$$
\begin{aligned}
\frac{da_{\mathrm{p,e}}}{dz} = i\frac{3}{16}kZ_0\chi^{(3)}_{xxxx}\Big\{&\left(J^*_{\mathrm{T1}}a^{*2}_{\mathrm{p,e}}a_{\mathrm{s,e}} + 2J^*_{\mathrm{T2}}a^*_{\mathrm{p,e}}a^*_{\mathrm{p,o}}a_{\mathrm{s,o}} + J^*_{\mathrm{T3}}a^{*2}_{\mathrm{p,o}}a_{\mathrm{s,e}}\right)e^{-i\Delta\beta z} \\
&+\left(J_{\mathrm{P1}}\left|a_{\mathrm{p,e}}\right|^2 a_{\mathrm{p,e}} + 2J_{\mathrm{P2}}\left|a_{\mathrm{p,o}}\right|^2 a_{\mathrm{p,e}} + J_{\mathrm{P3}}a^2_{\mathrm{p,o}}a^*_{\mathrm{p,e}}\right) \\
&+2\left(J_{\mathrm{X1}}\left|a_{\mathrm{s,e}}\right|^2 a_{\mathrm{p,e}} + J_{\mathrm{X2}}\left|a_{\mathrm{s,o}}\right|^2 a_{\mathrm{p,e}} + J_{\mathrm{F1}}a^*_{\mathrm{s,e}}a_{\mathrm{s,o}}a_{\mathrm{p,o}} + J_{\mathrm{F2}}a_{\mathrm{s,e}}a^*_{\mathrm{s,o}}a_{\mathrm{p,o}}\right)\Big\},
\end{aligned}
\tag{S10}
$$

$$\begin{aligned}\frac{da_{\text{p,o}}}{dz} = i\frac{3}{16}kZ_0\chi^{(3)}_{xxxx}\Big\{&\left(J^*_{\text{T4}}a^{*2}_{\text{p,o}}a_{\text{s,o}} + 2J^*_{\text{T3}}a^*_{\text{p,e}}a^*_{\text{p,o}}a_{\text{s,e}} + J^*_{\text{T2}}a^{*2}_{\text{p,e}}a_{\text{s,o}}\right)e^{-i\Delta\beta z}\\ &+\left(J_{\text{P4}}\left|a_{\text{p,o}}\right|^2 a_{\text{p,o}} + 2J_{\text{P2}}\left|a_{\text{p,e}}\right|^2 a_{\text{p,o}} + J^*_{\text{P3}}a^2_{\text{p,e}}a^*_{\text{p,o}}\right)\\ &+2\left(J_{\text{X3}}\left|a_{\text{s,e}}\right|^2 a_{\text{p,o}} + J_{\text{X4}}\left|a_{\text{s,o}}\right|^2 a_{\text{p,o}} + J^*_{\text{F1}}a_{\text{s,e}}a^*_{\text{s,o}}a_{\text{p,e}} + J^*_{\text{F2}}a^*_{\text{s,e}}a_{\text{s,o}}a_{\text{p,e}}\right)\Big\},\end{aligned} \tag{S11}$$

$$\begin{aligned}\frac{da_{\text{s,e}}}{dz} = i\frac{9}{16}kZ_0\chi^{(3)}_{xxxx}\Big\{&\left(\frac{1}{3}J_{\text{T1}}a^3_{\text{p,e}} + J_{\text{T3}}a_{\text{p,e}}a^2_{\text{p,o}}\right)e^{i\Delta\beta z}\\ &+\left(J_{\text{S1}}\left|a_{\text{s,e}}\right|^2 a_{\text{s,e}} + 2J_{\text{S2}}\left|a_{\text{s,o}}\right|^2 a_{\text{s,e}} + J_{\text{S3}}a^2_{\text{s,o}}a^*_{\text{s,e}}\right)\\ &+2\left(J_{\text{X1}}\left|a_{\text{p,e}}\right|^2 a_{\text{s,e}} + J_{\text{X3}}\left|a_{\text{p,o}}\right|^2 a_{\text{s,e}} + J_{\text{F1}}a^*_{\text{p,e}}a_{\text{p,o}}a_{\text{s,o}} + J^*_{\text{F2}}a_{\text{p,e}}a^*_{\text{p,o}}a_{\text{s,o}}\right)\Big\},\end{aligned} \tag{S12}$$

$$\begin{aligned}\frac{da_{\text{s,o}}}{dz} = i\frac{9}{16}kZ_0\chi^{(3)}_{xxxx}\Big\{&\left(\frac{1}{3}J_{\text{T4}}a^3_{\text{p,o}} + J_{\text{T2}}a^2_{\text{p,e}}a_{\text{p,o}}\right)e^{i\Delta\beta z}\\ &+\left(J_{\text{S4}}\left|a_{\text{s,o}}\right|^2 a_{\text{s,o}} + 2J_{\text{S2}}\left|a_{\text{s,e}}\right|^2 a_{\text{s,o}} + J^*_{\text{S3}}a^2_{\text{s,e}}a^*_{\text{s,o}}\right)\\ &+2\left(J_{\text{X2}}\left|a_{\text{p,e}}\right|^2 a_{\text{s,o}} + J_{\text{X4}}\left|a_{\text{p,o}}\right|^2 a_{\text{s,o}} + J^*_{\text{F1}}a_{\text{p,e}}a^*_{\text{p,o}}a_{\text{s,e}} + J_{\text{F2}}a^*_{\text{p,e}}a_{\text{p,o}}a_{\text{s,e}}\right)\Big\},\end{aligned} \tag{S13}$$

where $k = \omega/c$ is the pump wavenumber in a vacuum, and $\Delta\beta = 3\beta_{\text{p}} - \beta_{\text{s}}$ is the wavenumber mismatch for THG and degenerate TPDC.

$J$'s in Eqs. (S10)–(S13) are the overlap integrals representing the strengths of nonlinear optical processes involving the four interacting modes. $J_{\text{T}i}$ are the overlap integrals for THG and degenerate TPDC:

$$\begin{aligned}J_{\text{T1}} &= \int\left(\mathbf{e}_{\text{p,e}}\cdot\mathbf{e}_{\text{p,e}}\right)\left(\mathbf{e}_{\text{p,e}}\cdot\mathbf{e}^*_{\text{s,e}}\right)dA,\\ J_{\text{T2}} &= \frac{1}{3}\int\left[2\left(\mathbf{e}_{\text{p,e}}\cdot\mathbf{e}_{\text{p,o}}\right)\left(\mathbf{e}_{\text{p,e}}\cdot\mathbf{e}^*_{\text{s,o}}\right) + \left(\mathbf{e}_{\text{p,e}}\cdot\mathbf{e}_{\text{p,e}}\right)\left(\mathbf{e}_{\text{p,o}}\cdot\mathbf{e}^*_{\text{s,o}}\right)\right]dA,\\ J_{\text{T3}} &= \frac{1}{3}\int\left[2\left(\mathbf{e}_{\text{p,e}}\cdot\mathbf{e}_{\text{p,o}}\right)\left(\mathbf{e}_{\text{p,o}}\cdot\mathbf{e}^*_{\text{s,e}}\right) + \left(\mathbf{e}_{\text{p,o}}\cdot\mathbf{e}_{\text{p,o}}\right)\left(\mathbf{e}_{\text{p,e}}\cdot\mathbf{e}^*_{\text{s,e}}\right)\right]dA,\\ J_{\text{T4}} &= \int\left(\mathbf{e}_{\text{p,o}}\cdot\mathbf{e}_{\text{p,o}}\right)\left(\mathbf{e}_{\text{p,o}}\cdot\mathbf{e}^*_{\text{s,o}}\right)dA.\end{aligned} \tag{S14}$$

Here, $J_{\text{T1}}$ [$J_{\text{T4}}$] are for THG/TPDC driven by the even [odd] pump mode only, while $J_{\text{T2}}$ and $J_{\text{T3}}$ are for THG/TPDC excited by the pump field in a mixture of the even and odd modes.

$J_{\text{P}i}$ and $J_{\text{S}i}$ are for SPM of the pump and the third harmonic, respectively:

$$J_{\mathrm{P1}} = \frac{1}{3}\int\left[2\left|\mathbf{e}_{\mathrm{p,e}}\right|^4 + \left|\mathbf{e}_{\mathrm{p,e}}\cdot\mathbf{e}_{\mathrm{p,e}}\right|^2\right]dA,$$
$$J_{\mathrm{P2}} = \frac{1}{3}\int\left[\left|\mathbf{e}_{\mathrm{p,e}}\cdot\mathbf{e}_{\mathrm{p,o}}\right|^2 + \left|\mathbf{e}^*_{\mathrm{p,e}}\cdot\mathbf{e}_{\mathrm{p,o}}\right|^2 + \left|\mathbf{e}_{\mathrm{p,e}}\right|^2\left|\mathbf{e}_{\mathrm{p,o}}\right|^2\right]dA,$$
$$J_{\mathrm{P3}} = \frac{1}{3}\int\left[2\left(\mathbf{e}^*_{\mathrm{p,e}}\cdot\mathbf{e}_{\mathrm{p,o}}\right)^2 + \left(\mathbf{e}^*_{\mathrm{p,e}}\cdot\mathbf{e}^*_{\mathrm{p,e}}\right)\left(\mathbf{e}_{\mathrm{p,o}}\cdot\mathbf{e}_{\mathrm{p,o}}\right)\right]dA,$$
$$J_{\mathrm{P4}} = \frac{1}{3}\int\left[2\left|\mathbf{e}_{\mathrm{p,o}}\right|^4 + \left|\mathbf{e}_{\mathrm{p,o}}\cdot\mathbf{e}_{\mathrm{p,o}}\right|^2\right]dA. \tag{S15}$$

$$J_{\mathrm{S1}} = \frac{1}{3}\int\left[2\left|\mathbf{e}_{\mathrm{s,e}}\right|^4 + \left|\mathbf{e}_{\mathrm{s,e}}\cdot\mathbf{e}_{\mathrm{s,e}}\right|^2\right]dA,$$
$$J_{\mathrm{S2}} = \frac{1}{3}\int\left[\left|\mathbf{e}_{\mathrm{s,e}}\cdot\mathbf{e}_{\mathrm{s,o}}\right|^2 + \left|\mathbf{e}^*_{\mathrm{s,e}}\cdot\mathbf{e}_{\mathrm{s,o}}\right|^2 + \left|\mathbf{e}_{\mathrm{s,e}}\right|^2\left|\mathbf{e}_{\mathrm{s,o}}\right|^2\right]dA,$$
$$J_{\mathrm{S3}} = \frac{1}{3}\int\left[2\left(\mathbf{e}^*_{\mathrm{s,e}}\cdot\mathbf{e}_{\mathrm{s,o}}\right)^2 + \left(\mathbf{e}^*_{\mathrm{s,e}}\cdot\mathbf{e}^*_{\mathrm{s,e}}\right)\left(\mathbf{e}_{\mathrm{s,o}}\cdot\mathbf{e}_{\mathrm{s,o}}\right)\right]dA,$$
$$J_{\mathrm{S4}} = \frac{1}{3}\int\left[2\left|\mathbf{e}_{\mathrm{s,o}}\right|^4 + \left|\mathbf{e}_{\mathrm{s,o}}\cdot\mathbf{e}_{\mathrm{s,o}}\right|^2\right]dA. \tag{S16}$$

Here, $J_{\mathrm{P1}}$ [$J_{\mathrm{S1}}$] and $J_{\mathrm{P4}}$ [$J_{\mathrm{S4}}$] correspond to the SPM of the pump [third-harmonic] field in the even and the odd mode, respectively. On the other hand, $J_{\mathrm{P2}}$ [$J_{\mathrm{S2}}$] and $J_{\mathrm{P3}}$ [$J_{\mathrm{S3}}$] can be viewed as intermodal XPM and intermodal conversion, respectively, between the even and odd modes of the pump [third harmonic].

$J_{\mathrm{X}i}$ and $J_{\mathrm{F}i}$ are for XPM and FWM, respectively, between the pump and the third harmonic:

$$J_{\mathrm{X1}} = \frac{1}{3}\int\left[\left|\mathbf{e}_{\mathrm{p,e}}\cdot\mathbf{e}_{\mathrm{s,e}}\right|^2 + \left|\mathbf{e}_{\mathrm{p,e}}\cdot\mathbf{e}^*_{\mathrm{s,e}}\right|^2 + \left|\mathbf{e}_{\mathrm{p,e}}\right|^2\left|\mathbf{e}_{\mathrm{s,e}}\right|^2\right]dA,$$
$$J_{\mathrm{X2}} = \frac{1}{3}\int\left[\left|\mathbf{e}_{\mathrm{p,e}}\cdot\mathbf{e}_{\mathrm{s,o}}\right|^2 + \left|\mathbf{e}_{\mathrm{p,e}}\cdot\mathbf{e}^*_{\mathrm{s,o}}\right|^2 + \left|\mathbf{e}_{\mathrm{p,e}}\right|^2\left|\mathbf{e}_{\mathrm{s,o}}\right|^2\right]dA,$$
$$J_{\mathrm{X3}} = \frac{1}{3}\int\left[\left|\mathbf{e}_{\mathrm{p,o}}\cdot\mathbf{e}_{\mathrm{s,e}}\right|^2 + \left|\mathbf{e}_{\mathrm{p,o}}\cdot\mathbf{e}^*_{\mathrm{s,e}}\right|^2 + \left|\mathbf{e}_{\mathrm{p,o}}\right|^2\left|\mathbf{e}_{\mathrm{s,e}}\right|^2\right]dA,$$
$$J_{\mathrm{X4}} = \frac{1}{3}\int\left[\left|\mathbf{e}_{\mathrm{p,o}}\cdot\mathbf{e}_{\mathrm{s,o}}\right|^2 + \left|\mathbf{e}_{\mathrm{p,o}}\cdot\mathbf{e}^*_{\mathrm{s,o}}\right|^2 + \left|\mathbf{e}_{\mathrm{p,o}}\right|^2\left|\mathbf{e}_{\mathrm{s,o}}\right|^2\right]dA. \tag{S17}$$

$$J_{\mathrm{F1}} = \frac{1}{3}\int\left[\left(\mathbf{e}^*_{\mathrm{p,e}}\cdot\mathbf{e}_{\mathrm{p,o}}\right)\left(\mathbf{e}^*_{\mathrm{s,e}}\cdot\mathbf{e}_{\mathrm{s,o}}\right) + \left(\mathbf{e}^*_{\mathrm{p,e}}\cdot\mathbf{e}_{\mathrm{s,o}}\right)\left(\mathbf{e}_{\mathrm{p,o}}\cdot\mathbf{e}^*_{\mathrm{s,e}}\right) + \left(\mathbf{e}^*_{\mathrm{p,e}}\cdot\mathbf{e}^*_{\mathrm{s,e}}\right)\left(\mathbf{e}_{\mathrm{p,o}}\cdot\mathbf{e}_{\mathrm{s,o}}\right)\right]dA,$$
$$J_{\mathrm{F2}} = \frac{1}{3}\int\left[\left(\mathbf{e}^*_{\mathrm{p,e}}\cdot\mathbf{e}_{\mathrm{p,o}}\right)\left(\mathbf{e}_{\mathrm{s,e}}\cdot\mathbf{e}^*_{\mathrm{s,o}}\right) + \left(\mathbf{e}^*_{\mathrm{p,e}}\cdot\mathbf{e}^*_{\mathrm{s,o}}\right)\left(\mathbf{e}_{\mathrm{p,o}}\cdot\mathbf{e}_{\mathrm{s,e}}\right) + \left(\mathbf{e}^*_{\mathrm{p,e}}\cdot\mathbf{e}_{\mathrm{s,e}}\right)\left(\mathbf{e}_{\mathrm{p,o}}\cdot\mathbf{e}^*_{\mathrm{s,o}}\right)\right]dA. \tag{S18}$$

In Eq. (S17), $J_{\mathrm{X1}}$ [$J_{\mathrm{X4}}$] is for XPM between the even [odd] pump mode and the even [odd] third-harmonic mode. On the other hand, $J_{\mathrm{X2}}$ [$J_{\mathrm{X3}}$] is for XPM between the even [odd] pump mode and the odd [even] third-harmonic mode. In Eq. (S18), $J_{\mathrm{F1}}$ and $J_{\mathrm{F2}}$ are for FWM among four modes (2 for the pump and 2 for the third harmonic). $J_{\mathrm{F1}}$ causes intermodal conversion in the same direction (either even to odd or odd to even) for the pump and the third harmonic, whereas $J_{\mathrm{F2}}$ in the opposite direction. To appreciate the physical meanings of all the overlap integrals graphically, we illustrate the energy-level diagram for the third-order nonlinear optical process corresponding to each overlap integral in Fig. S2.

We also numerically calculate the overlap integrals for three different third-harmonic hybrid modes, the $HE_{12}$, the $EH_{11}$, and the $HE_{31}$ modes, in silica glass optical nanofibers with respective phase-matching diameters, as summarized in Supplementary Table 1. For optical waveguides of perfectly circular cross-sections that preserve the degeneracy of the even and odd hybrid modes, the overlap integrals for SPM have the following symmetries.

$$\begin{aligned} J_{\mathrm{P1}} &= J_{\mathrm{P4}}, \\ J_{\mathrm{S1}} &= J_{\mathrm{S4}}. \end{aligned} \tag{S19}$$

For XPM/FWM and THG/TPDC, the symmetry properties of the overlap integrals depend on the azimuthal number of the third-harmonic hybrid mode. For third-harmonic modes having the azimuthal number 1 (e.g., the $HE_{12}$ and the $EH_{11}$ modes),

$$\begin{aligned} J_{\mathrm{T1}} = 3J_{\mathrm{T2}} = 3J_{\mathrm{T3}} &= J_{\mathrm{T4}}, \\ J_{\mathrm{X1}} &= J_{\mathrm{X4}}, \\ J_{\mathrm{X2}} &= J_{\mathrm{X3}}, \end{aligned} \tag{S20}$$

whereas for third-harmonic modes with the azimuthal number 3 (e.g., the $HE_{31}$ mode),

$$\begin{aligned} J_{\mathrm{T1}} = J_{\mathrm{T2}} = -J_{\mathrm{T3}} &= -J_{\mathrm{T4}}, \\ J_{\mathrm{X1}} = J_{\mathrm{X2}} = J_{\mathrm{X3}} &= J_{\mathrm{X4}}, \\ J_{\mathrm{F1}} &= -J_{\mathrm{F2}}. \end{aligned} \tag{S21}$$

We note that the symmetry relations among $J_{\mathrm{T}i}$'s in Eqs. (S20) and (S21) are closely related to the total angular momentum (TAM) conservation in the THG process, which we will discuss in Supplementary Note 3.

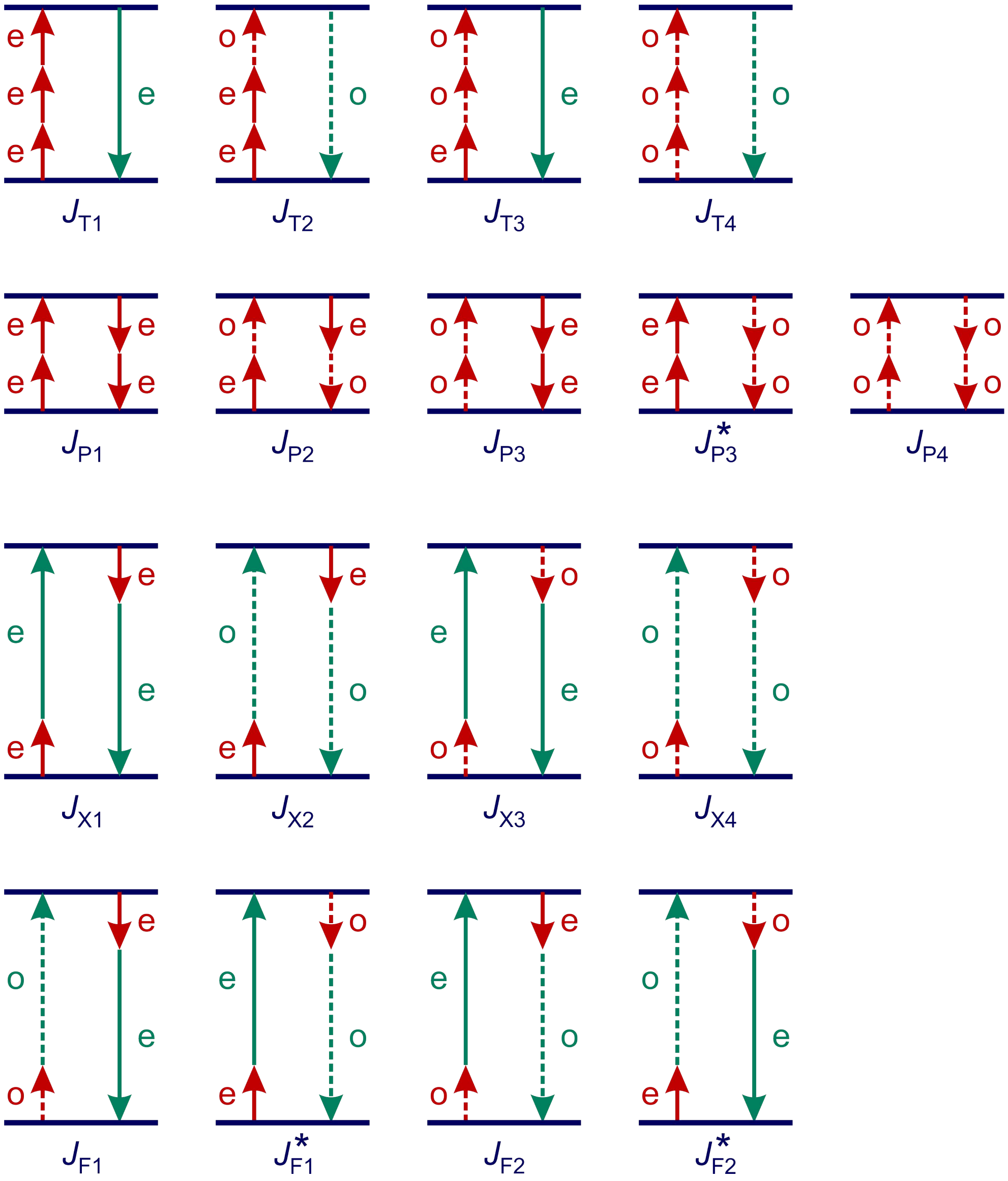

**FIG. S2. Energy level diagrams for third-order nonlinear optical processes involving hybrid modes in an optical nanofiber.** The red and turquoise represent the pump and third-harmonic photons, and the solid and dashed arrows correspond to the even ('e') and odd ('o') hybrid modes. Each overlap integral $J$ is defined as Eqs. (S14)–(S18). The diagrams for the self-phase modulation of the third harmonic ($J_{\mathrm{S}i}$) have the same configurations as those of the pump ($J_{\mathrm{P}i}$) ($i = 1$ to $4$). The diagrams for $J^*$ can be obtained by inverting or 180º-rotating those for $J$. Note that $J^* = J$ because $J$ is real-valued.

**Supplementary Table 1. Calculated overlap integrals for third-order nonlinear optical processes in silica glass optical nanofibers driven by the pump beam in the $HE_{11}$ mode at 1550 nm wavelength. $J_{T,OVM}$ is defined as Eq. (S28) in Supplementary Note 3.**

| | | | | | |
|---|---|---|---|---|---|
| Third-harmonic modes | Hybrid mode | | $HE_{12}$ | $EH_{11}$ | $HE_{31}$ |
| | Optical vortex mode | | $OV^{\pm}_{0,2}$ | $OV^{\mp}_{\pm2,1}$ | $OV^{\pm}_{\pm2,1}$ |
| | Total angular momentum | | ±1 | ±1 | ±3 |
| Phase-matching nanofiber diameter (nm) | | | 762 | 624 | 717 |
| Overlap integrals ($\mu m^{-2}$) | THG and degenerate TPDC | $J_{T1}$ | 0.3703 | 0.0043 | 0.0767 |
| | | $J_{T2}$ | 0.1234 | 0.0014 | 0.0767 |
| | | $J_{T3}$ | 0.1234 | 0.0014 | -0.0767 |
| | | $J_{T4}$ | 0.3703 | 0.0043 | -0.0767 |
| | | $J_{T,OVM}$ | 0 | 0 | 0.1534 |
| | SPM of the pump | $J_{P1}$ | 0.9642 | 0.3332 | 0.7516 |
| | | $J_{P2}$ | 0.3635 | 0.1243 | 0.2830 |
| | | $J_{P3}$ | 0.2372 | 0.0845 | 0.1855 |
| | | $J_{P4}$ | 0.9642 | 0.3332 | 0.7516 |
| | SPM of the third harmonic | $J_{S1}$ | 3.8831 | 4.3102 | 4.0835 |
| | | $J_{S2}$ | 1.5385 | 1.5774 | 1.7437 |
| | | $J_{S3}$ | 0.8062 | 1.1553 | 0.5961 |
| | | $J_{S4}$ | 3.8831 | 4.3102 | 4.0835 |
| | XPM | $J_{X1}$ | 1.4354 | 0.7598 | 1.0060 |
| | | $J_{X2}$ | 0.6440 | 0.6528 | 1.0060 |
| | | $J_{X3}$ | 0.6440 | 0.6528 | 1.0060 |
| | | $J_{X4}$ | 1.4354 | 0.7598 | 1.0060 |
| | FWM | $J_{F1}$ | 0.3499 | 0.0020 | -0.1584 |
| | | $J_{F2}$ | 0.4415 | 0.1050 | 0.1584 |

**Supplementary Note 3: Role of the spin-orbit coupling (SOC) and the resulting spin-orbit entanglement (SOE) in third-harmonic optical vortex generation (TH-OVG)**

In the previous analysis in Supplementary Note 2, the electric fields are expressed in the basis of the even and odd hybrid modes. Here, we rewrite the nonlinear coupled-mode equations in Eqs. (S10)–(S13) and the overlap integrals $J$'s in Eqs. (S14) –(S18) in the OVM basis to elucidate the role of SOC and SOE in TH-OVG. The normalized field distribution of each OVM is related to those of the constituent even and odd hybrid modes as below.

$$\begin{aligned}\mathbf{e}_{\mathrm{p},\pm} &= \frac{1}{\sqrt{2}}\left(\mathbf{e}_{\mathrm{p,e}} \pm i\mathbf{e}_{\mathrm{p,o}}\right),\\ \mathbf{e}_{\mathrm{s},\pm} &= \frac{1}{\sqrt{2}}\left(\mathbf{e}_{\mathrm{s,e}} \pm i\mathbf{e}_{\mathrm{s,o}}\right).\end{aligned} \tag{S22}$$

Considering this relationship, we express the slowly varying complex field amplitude of each OVM as follows.

$$\begin{aligned}a_{\mathrm{p},\pm} &= \frac{1}{\sqrt{2}}\left(a_{\mathrm{p,e}} \mp i a_{\mathrm{p,o}}\right),\\ a_{\mathrm{s},\pm} &= \frac{1}{\sqrt{2}}\left(a_{\mathrm{s,e}} \mp i a_{\mathrm{s,o}}\right).\end{aligned} \tag{S23}$$

Eqs. (S10) and (S11) can then be combined into

$$\frac{da_{\mathrm{p},\pm}}{dz} = i\frac{3}{16}kZ_0\chi^{(3)}_{xxxx}\frac{1}{4}\begin{bmatrix}\left(J_{\mathrm{P1}} + 4J_{\mathrm{P2}} - 2J_{\mathrm{P3}} + J_{\mathrm{P4}}\right)\left|a_{\mathrm{p},\pm}\right|^2 a_{\mathrm{p},\pm}\\ +2\left(J_{\mathrm{P1}} + 2J_{\mathrm{P3}} + J_{\mathrm{P4}}\right)\left|a_{\mathrm{p},\mp}\right|^2 a_{\mathrm{p},\pm}\\ +\left(J_{\mathrm{P1}} - 4J_{\mathrm{P2}} - 2J_{\mathrm{P3}} + J_{\mathrm{P4}}\right)a^2_{\mathrm{p},\mp}a^*_{\mathrm{p},\pm}\\ +\left(J_{\mathrm{P1}} - J_{\mathrm{P4}}\right)\left(\left|a_{\mathrm{p},\mp}\right|^2 a_{\mathrm{p},\mp} + 2\left|a_{\mathrm{p},\pm}\right|^2 a_{\mathrm{p},\mp} + a^2_{\mathrm{p},\pm}a^*_{\mathrm{p},\mp}\right)\end{bmatrix}, \tag{S24}$$

whereas Eqs. (S12) and (S13) into

$$\frac{da_{\mathrm{s},\pm}}{dz} = i\frac{9}{16}kZ_0\chi^{(3)}_{xxxx}\frac{1}{4}\left[\begin{array}{l}\left\{\begin{array}{l}\frac{1}{3}\left(J_{\mathrm{T1}}+3J_{\mathrm{T2}}-3J_{\mathrm{T3}}-J_{\mathrm{T4}}\right)a^3_{\mathrm{p},\pm} \\ +\left(J_{\mathrm{T1}}+J_{\mathrm{T2}}+J_{\mathrm{T3}}+J_{\mathrm{T4}}\right)a^2_{\mathrm{p},\pm}a_{\mathrm{p},\mp} \\ +\left(J_{\mathrm{T1}}-J_{\mathrm{T2}}+J_{\mathrm{T3}}-J_{\mathrm{T4}}\right)a_{\mathrm{p},\pm}a^2_{\mathrm{p},\mp} \\ +\frac{1}{3}\left(J_{\mathrm{T1}}-3J_{\mathrm{T2}}-3J_{\mathrm{T3}}+J_{\mathrm{T4}}\right)a^3_{\mathrm{p},\mp}\end{array}\right\}e^{i\Delta\beta z} \\ +2\left\{\begin{array}{l}\left(J_{\mathrm{X1}}+J_{\mathrm{X2}}+J_{\mathrm{X3}}+J_{\mathrm{X4}}\right)\left(\left|a_{\mathrm{p},+}\right|^2+\left|a_{\mathrm{p},-}\right|^2\right)a_{\mathrm{s},\pm} \\ +\left(J_{\mathrm{X1}}+J_{\mathrm{X2}}-J_{\mathrm{X3}}-J_{\mathrm{X4}}\right)\left(a_{\mathrm{p},+}a^*_{\mathrm{p},-}+a^*_{\mathrm{p},+}a_{\mathrm{p},-}\right)a_{\mathrm{s},\pm} \\ +\left(J_{\mathrm{X1}}-J_{\mathrm{X2}}-J_{\mathrm{X3}}+J_{\mathrm{X4}}\right)\left(a_{\mathrm{p},+}a^*_{\mathrm{p},-}+a^*_{\mathrm{p},+}a_{\mathrm{p},-}\right)a_{\mathrm{s},\mp} \\ +\left(J_{\mathrm{X1}}-J_{\mathrm{X2}}+J_{\mathrm{X3}}-J_{\mathrm{X4}}\right)\left(\left|a_{\mathrm{p},+}\right|^2+\left|a_{\mathrm{p},-}\right|^2\right)a_{\mathrm{s},\mp} \\ \pm2\left(-J_{\mathrm{F1}}+J_{\mathrm{F2}}\right)\left(\left|a_{\mathrm{p},+}\right|^2-\left|a_{\mathrm{p},-}\right|^2\right)a_{\mathrm{s},\pm} \\ \pm2\left(J_{\mathrm{F1}}+J_{\mathrm{F2}}\right)\left(a_{\mathrm{p},+}a^*_{\mathrm{p},-}-a^*_{\mathrm{p},+}a_{\mathrm{p},-}\right)a_{\mathrm{s},\mp}\end{array}\right\}\end{array}\right], \tag{S25}$$

where we use the undepleted-pump approximation [5] and assume $\left|a_{\mathrm{s}}\right| \ll \left|a_{\mathrm{p}}\right|$ so as to take into account only THG, SPM of the pump, and XPM of the third harmonic combined with FWM driven by the pump. In Eqs. (S24) and (S25), several terms vanish due to the symmetry relations among the overlap integrals in Eqs. (S19)–(S21). In particular, since $J_{\mathrm{P1}} - J_{\mathrm{P4}} = 0$ according to Eq. (S19), the last term in Eq. (S24) is zero, so SPM-induced conversion between the two pump OVMs of mutually opposite spins is inhibited. Moreover, in Eq. (S25), the last THG term and the last XPM term also vanish because $J_{\mathrm{T1}} - 3J_{\mathrm{T2}} - 3J_{\mathrm{T3}} + J_{\mathrm{T4}} = 0$ and $J_{\mathrm{X1}} - J_{\mathrm{X2}} + J_{\mathrm{X3}} - J_{\mathrm{X4}} = 0$ owing to the relations in Eqs. (S20) and (S21), which indicates that the pump in one OVM cannot create the third harmonic in another of the opposite spin. Equation (S25) suggests that a pure third-harmonic OVM can be generated by the pump in a single OVM, i.e., when either $a_{\mathrm{p},+}$ or $a_{\mathrm{p},-}$ is zero. In this scenario, Eqs. (S24) and (S25) are simplified as

$$\frac{da_{\mathrm{p},\pm}}{dz} = i\frac{3}{16}kZ_0\chi^{(3)}_{xxxx}J_{\mathrm{P,OVM}}\left|a_{\mathrm{p},\pm}\right|^2 a_{\mathrm{p},\pm}, \tag{S26}$$

$$\frac{da_{\mathrm{s},\pm}}{dz} = i\frac{9}{16}kZ_0\chi^{(3)}_{xxxx}\left(\frac{1}{3}J_{\mathrm{T,OVM}}a^3_{\mathrm{p},\pm}e^{i\Delta\beta z} + 2J_{\mathrm{X,OVM}}\left|a_{\mathrm{p},\pm}\right|^2 a_{\mathrm{s},\pm}\right), \tag{S27}$$

$$J_{\mathrm{T,OVM}} = \frac{1}{4}\left(J_{\mathrm{T1}} + 3J_{\mathrm{T2}} - 3J_{\mathrm{T3}} - J_{\mathrm{T4}}\right) \\ = 2\int\left[\left(\mathbf{e}_{\mathrm{p},+}\cdot\mathbf{e}_{\mathrm{p},+}\right)\left(\mathbf{e}_{\mathrm{p},+}\cdot\mathbf{e}_{\mathrm{s},+}^{*}\right) + \left(\mathbf{e}_{\mathrm{p},-}\cdot\mathbf{e}_{\mathrm{p},-}\right)\left(\mathbf{e}_{\mathrm{p},-}\cdot\mathbf{e}_{\mathrm{s},-}^{*}\right)\right]dA. \tag{S28}$$

$$J_{\mathrm{P,OVM}} = \frac{1}{4}\left(J_{\mathrm{P1}} + 4J_{\mathrm{P2}} - 2J_{\mathrm{P3}} + J_{\mathrm{P4}}\right), \\ J_{\mathrm{X,OVM}} = \frac{1}{4}\left[J_{\mathrm{X1}} + J_{\mathrm{X2}} + J_{\mathrm{X3}} + J_{\mathrm{X4}} + 2\left(-J_{\mathrm{F1}} + J_{\mathrm{F2}}\right)\right]. \tag{S29}$$

For the $\mathrm{OV}_{0,2}^{\pm}$ and the $\mathrm{OV}_{\pm 2,1}^{\mp}$ third-harmonic modes, $J_{\mathrm{T,OVM}}$ vanishes because of the symmetry relations of $J_{\mathrm{T}i}$'s in Eq. (S20), so the THG in those OVMs by the pump in the $\mathrm{OV}_{0,1}^{\pm}$ mode is forbidden. In strong contrast, $J_{\mathrm{T,OVM}}$ is nonzero for the $\mathrm{OV}_{\pm 2,1}^{\pm}$ third-harmonic mode according to Eq. (S21), so the TH-OVG is permitted.

These results are also anticipated from the TAM conservation. We emphasize that the nonzero $J_{\mathrm{T,OVM}}$ is facilitated by SOC that releases the requirement of the simultaneous conservation of the OAM and SAM. The normalized field distribution of the OVM, $\mathrm{OV}_{l,m}^{\pm}$, in Eq. (S22) may be expressed in an alternative form of [4,6]

$$\mathbf{e}_{\pm} \equiv u_{\pm,\mathrm{L}}(r,\phi)\hat{\boldsymbol{\sigma}}_{+} + u_{\pm,\mathrm{R}}(r,\phi)\hat{\boldsymbol{\sigma}}_{-} + u_{\pm,z}(r,\phi)\hat{\mathbf{z}}, \tag{S30}$$

$$\begin{bmatrix} u_{+,\mathrm{L}} \\ u_{+,\mathrm{R}} \\ u_{+,z} \end{bmatrix} \equiv \begin{bmatrix} iA_{j,m}(r)e^{i(j-1)\phi} \\ iB_{j,m}(r)e^{i(j+1)\phi} \\ C_{j,m}(r)e^{ij\phi} \end{bmatrix}, \tag{S31}$$

$$\begin{bmatrix} u_{-,\mathrm{L}} \\ u_{-,\mathrm{R}} \\ u_{-,z} \end{bmatrix} \equiv \begin{bmatrix} iB_{j,m}(r)e^{-i(j+1)\phi} \\ iA_{j,m}(r)e^{-i(j-1)\phi} \\ C_{j,m}(r)e^{-ij\phi} \end{bmatrix}, \tag{S32}$$

where $\hat{\boldsymbol{\sigma}}_{\pm} = (\hat{\mathbf{x}} \pm i\hat{\mathbf{y}})/\sqrt{2}$ are the transverse unit vectors along the left-handed ($\hat{\boldsymbol{\sigma}}_{+}$) and right-handed ($\hat{\boldsymbol{\sigma}}_{-}$) circular polarizations, whereas $\hat{\mathbf{z}}$ is the axial unit vector, and $(r,\phi,z)$ are the cylindrical coordinates. $l$ and $j = l \pm 1$ are the canonical OAM and TAM numbers of the modes, respectively, in the paraxial regime, where $\sigma = \pm 1$ is the canonical SAM number. We note that $j$ is the same as the azimuthal mode number of the hybrid mode that composes the OVM. In the presence of SOC, all the three real-valued functions, $A_{j,m}(r)$, $B_{j,m}(r)$, and $C_{j,m}(r)$, are non-negligible, as can be seen in Fig. S3. Then, the OVM in Eq. (S30) becomes a spin-orbit entangled state, its orbital and spin degrees of freedom being coupled to each other. The polarization gets deviated from perfectly circular (Fig. S1(d)), and the canonical OAM and SAM are not integers [7], i.e., the OVM is no longer an eigenmode of either the OAM or SAM, while it is still a TAM eigenmode with an integer TAM value, as shown in Fig. S4. In this case, by inserting Eqs. (S30)–(S32) into Eq. (S28), we find that the azimuthal integral for $J_{\mathrm{T,OVM}}$ takes the form of

$$J_{\mathrm{T,OVM}} \propto \int d\phi \cos\Big[\Big(j_{\mathrm{s}} - 3j_{\mathrm{p}}\Big)\phi\Big], \tag{S33}$$

where $j_{\mathrm{p}}$ and $j_{\mathrm{s}}$ are the TAM numbers of the pump and the third-harmonic OVM, respectively. This result indicates that $J_{\mathrm{T,OVM}}$ is nonzero, and thus TH-OVG can take place, only if the TAM can be conserved, i.e., $\Delta j \equiv j_{\mathrm{s}} - 3j_{\mathrm{p}} = 0$. Otherwise, TH-OVG is forbidden. On the other hand, the simultaneous conservation of the OAM and SAM is no longer required. When the pump is in the fundamental $\mathrm{OV}_{0,1}^{\pm}$ mode ($j_{\mathrm{p}} = \pm 1$), $J_{\mathrm{T,OVM}} = 0$ for THG in the $\mathrm{OV}_{0,2}^{\pm}$ and the $\mathrm{OV}_{\pm 2,1}^{\mp}$ mode with $j_{\mathrm{s}} = \pm 1$ and $\Delta j \neq 0$, whereas $J_{\mathrm{T,OVM}}$ is nonzero for THG in the $\mathrm{OV}_{\pm 2,1}^{\pm}$ mode having $j_{\mathrm{s}} = \pm 3$ and $\Delta j = 0$. These results were also obtained from the previous analysis of $J_{\mathrm{T,OVM}}$ using the symmetry relations in Eqs. (S20) and (S21), as shown in Supplementary Table 1.

In strong contrast, light propagating in the paraxial or weakly guiding regime exhibits radically different behaviors of TH-OVG. In general, the wave equation for the transverse ($\mathbf{e}_{\perp}$) and longitudinal ($\mathbf{e}_{\mathrm{z}}$) electric field components of a guided mode in an optical waveguide can be derived from Maxwell’s equations [4]:

$$\Big\{\nabla_{\perp}^{2} + k_{0}{}^{2} n^{2}(x,y) - \beta^{2}\Big\}\begin{bmatrix}\mathbf{e}_{\perp}\\ \mathbf{e}_{\mathrm{z}}\end{bmatrix} = -\begin{bmatrix}\nabla_{\perp}\\ i\beta\hat{\mathbf{z}}\end{bmatrix}\Big\{\mathbf{e}_{\perp}\cdot\nabla_{\perp}(\ln n^{2})\Big\}, \tag{S34}$$

where $k_{0}$ and $n(x,y)$ are the wavevector in a vacuum and the refractive index profile of the waveguide, respectively, and $\beta$ is the propagation constant of the mode. $\nabla_{\perp}$ is the transverse gradient. For weakly guiding optical waveguides with the tiny refractive index difference between the core and cladding, the right-hand side of Eq. (S34) can be neglected:

$$\begin{gathered}\left\{\nabla_\perp^2 + k^2 n^2(x,y) - \tilde{\beta}^2\right\}\tilde{\mathbf{e}}_\perp \simeq 0,\\ \tilde{\mathbf{e}}_z \simeq 0.\end{gathered} \tag{S35}$$

In this case, the electric field distribution of an OVM in Eqs. (S30)–(S32) can be simplified, as can be noticed from Fig. S3. For the spin-orbit aligned OVMs composed of the HE hybrid modes, $B_{j,m}(r)$ and $C_{j,m}(r)$ become negligible, so $\tilde{\mathbf{e}}_\pm \simeq i\tilde{A}_{j,m}(r)e^{\pm il\phi}\hat{\boldsymbol{\sigma}}_\pm$. On the other hand, for the spin-orbit anti-aligned OVMs consisting of the EH hybrid modes, $A_{j,m}(r)$ and $C_{j,m}(r)$ are negligible, so $\tilde{\mathbf{e}}_\pm \simeq i\tilde{B}_{j,m}(r)e^{\pm il\phi}\hat{\boldsymbol{\sigma}}_\mp$ [4,6]. Consequently, the SOC disappears, and the OVM becomes a product state rather than a spin-orbit entangled state, i.e., its orbital and spin degrees of freedom are completely separated from each other. The polarization is perfectly circular, and the canonical OAM and SAM have integer values, i.e., the OVM becomes a simultaneous eigenmode of the OAM and SAM, as well as the TAM. The azimuthal integral for $J_{\mathrm{T,OVM}}$ in Eq. (S28) can then be expressed as

$$\begin{aligned}\tilde{J}_{\mathrm{T,OVM}} &= 2\int\left[\left(\tilde{\mathbf{e}}_{\mathrm{p},+}\cdot\tilde{\mathbf{e}}_{\mathrm{p},+}\right)\left(\tilde{\mathbf{e}}_{\mathrm{p},+}\cdot\tilde{\mathbf{e}}^*_{\mathrm{s},+}\right) + \left(\tilde{\mathbf{e}}_{\mathrm{p},-}\cdot\tilde{\mathbf{e}}_{\mathrm{p},-}\right)\left(\tilde{\mathbf{e}}_{\mathrm{p},-}\cdot\tilde{\mathbf{e}}^*_{\mathrm{s},-}\right)\right]dA\\ &\propto \delta_{\sigma_\mathrm{s},3\sigma_\mathrm{p}}\int d\phi\cos\left[\left(l_\mathrm{s} - 3l_\mathrm{p}\right)\phi\right] = 0.\end{aligned} \tag{S36}$$

Here, $\delta$ is the Kronecker delta. Eq. (S36) implies that $\tilde{J}_{\mathrm{T,OVM}}$ is nonzero only when both the OAM and SAM are conserved individually during the THG process, which is, however, impossible because the SAM conservation cannot be achieved with $\sigma_{\mathrm{p,s}} = \pm 1$ $(\sigma_\mathrm{s} \neq 3\sigma_\mathrm{p})$. Therefore, TH-OVG is forbidden without SOC, e.g., in the paraxial regime and under the weakly guiding approximation.

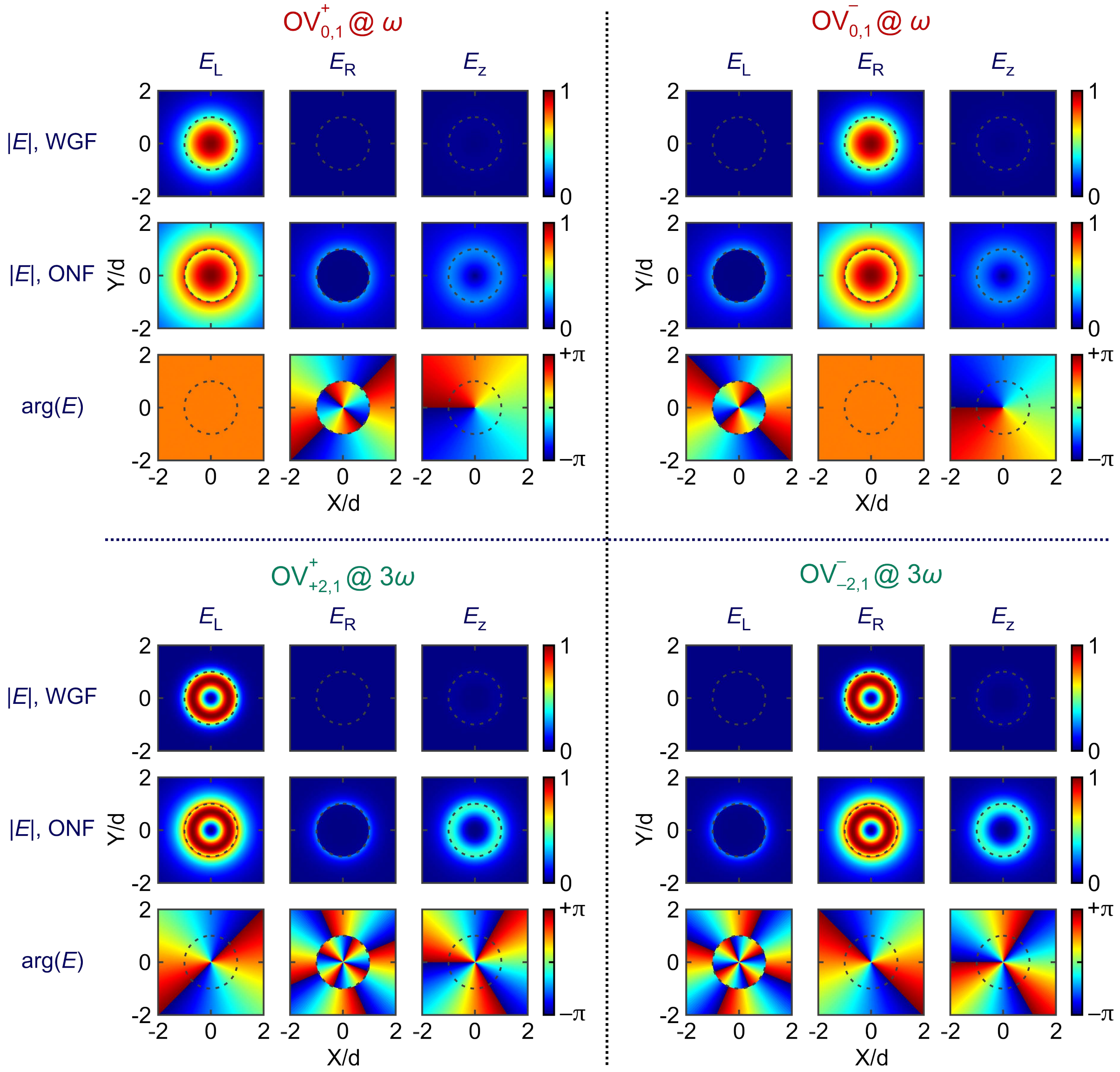

**FIG. S3. Comparison of the electric field components of optical vortex modes (OVMs) between a weakly guiding fiber (WGF) and an optical nanofiber (ONF).** We consider four OVMs in Fig. S1, the $OV_{0,1}^{\pm}$ mode at 1550 nm wavelength and the $OV_{\pm2,1}^{\pm}$ mode at the third harmonic. The WGF is a standard step-index fiber having a core diameter of 8.7 μm and a numerical aperture of 0.13. The ONF has a diameter of 717 nm. $E_L$ and $E_R$ are the left-handed and right-handed circularly polarized field components, whereas $E_z$ is the longitudinal one. The grey dashed circles indicate the core-cladding boundary for the WGF ($2d$ = 8.7 μm) and the air-silica interface for the ONF ($2d$ = 717 nm). The lowermost row for each OVM stands for the phase profile of each field component relative to the phase of $E_z$ on the +x-axis, which is identical for the WGF and ONF.

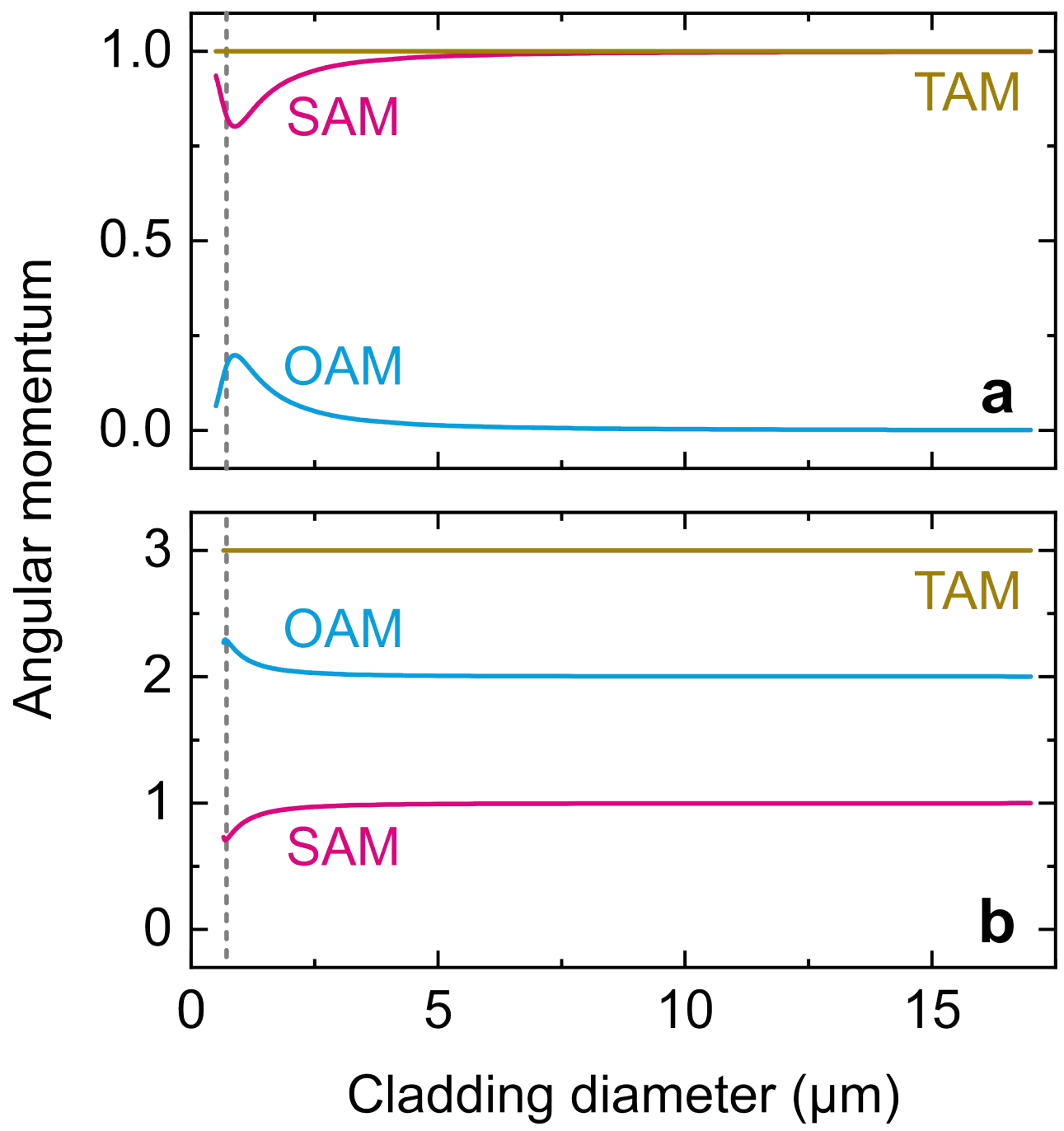

**FIG. S4. Calculated canonical angular momenta of optical vortex modes (OVMs) according to the cladding diameter in the course of tapering of an etched silica fiber.** (**a**, **b**) Orbital angular momentum (OAM, cyan), spin angular momentum (SAM, magenta), and total angular momentum (TAM, dark yellow) of the $\mathrm{OV}^{+}_{0,1}$ mode at the 1550 nm pump wavelength (**a**) and the $\mathrm{OV}^{+}_{+2,1}$ mode at the third harmonic (**b**), which interact through third-harmonic optical vortex generation (TH-OVG). As the cladding diameter reduces to the strongly guiding regime, the canonical OAM and SAM deviate from the integer values for the paraxial regime, while the TAM is conserved as an integer regardless of the cladding diameter variation. For the $\mathrm{OV}^{-}_{0,1}$ and the $\mathrm{OV}^{-}_{-2,1}$ modes, all the OAM, SAM, and TAM have the opposite signs and the same magnitudes as those of the $\mathrm{OV}^{+}_{0,1}$ and the $\mathrm{OV}^{+}_{+2,1}$ modes. The initial core [cladding] diameter and numerical aperture of the cladding-etched fiber are set as the experimental values of 8.7 [17] μm and 0.13, respectively, and the grey vertical dashed lines indicate the phase-matching nanofiber diameter of 717 nm for the TH-OVG process.

**Supplementary Note 4: Effect of birefringence on third-harmonic optical vortex generation (TH-OVG)**

When the cross-section of an optical nanofiber deviates from the perfectly circular one, the degeneracy between the hybrid even and odd modes can be broken. Then, TH-OVG can be viewed as a coherent combination of four distinct THG processes of different wavenumber mismatches. In this case, the THG part in the differential equation in Eq. (S27) is modified as follows:

$$\frac{da_{\mathrm{s},\pm}}{dz} = i\frac{9}{16}kZ_0\chi^{(3)}_{xxxx}\frac{1}{4}\left(\frac{1}{3}J_{\mathrm{T1}}e^{i\Delta\beta_1 z} + J_{\mathrm{T2}}e^{i\Delta\beta_2 z} - J_{\mathrm{T3}}e^{i\Delta\beta_3 z} - \frac{1}{3}J_{\mathrm{T4}}e^{i\Delta\beta_4 z}\right)a^3_{\mathrm{p},\pm}, \qquad \text{(S37)}$$

where $\Delta\beta_i$ ($i$ = 1 to 4) is the wavenumber mismatch for each THG process defined as

$$\begin{aligned} \Delta\beta_1 &= 3\beta_{\mathrm{p,e}} - \beta_{\mathrm{s,e}}, \\ \Delta\beta_2 &= 2\beta_{\mathrm{p,e}} + \beta_{\mathrm{p,o}} - \beta_{\mathrm{s,o}}, \\ \Delta\beta_3 &= \beta_{\mathrm{p,e}} + 2\beta_{\mathrm{p,o}} - \beta_{\mathrm{s,e}}, \\ \Delta\beta_4 &= 3\beta_{\mathrm{p,o}} - \beta_{\mathrm{s,o}}. \end{aligned} \qquad \text{(S38)}$$

In our case, such an imperfection of the nanofiber cross-section can arise during the deep wet-etching step of the nanofiber fabrication process. We observe the cross-section of deep cladding-wet-etched fiber with a scanning electron microscope (SEM) while one end is polished with focused ion beam (FIB) milling. Since the sample orientation with respect to the electron detector in the FIB-SEM system is difficult to identify precisely, it is almost impossible to directly determine the fiber cladding geometry. Instead, assuming that the germanium-doped fiber core has a perfectly circular cross-section, we compare the eccentricity between the cladding and the core that appear in the scanning electron micrograph to determine the cladding eccentricity (0.05 in the case of Supplementary Fig. S5). Here, the deformation is approximated to be elliptical, and its eccentricity is defined as $\sqrt{1-\left(b/a\right)^2}$ , where $a$ and $b$ are the major and minor axes, respectively, of the elliptical cross-section.

The phase-matching conditions for TH-OVG depend on the deformation of the nanofiber cross-section, as the four constituent THG processes are generally phase matched at different nanofiber diameters according to Eq. (S38). For efficient TH-OVG with high suppression of THG in the unwanted $HE_{12}$ mode, the pump polarization should be maintained circular along the nanofiber, which is, however, hard to achieve in the presence of the deformation-induced nanofiber birefringence. Then, THG in the $HE_{12}$ mode cannot be fully suppressed with input pump polarization control. Furthermore, the degeneracy between the third-harmonic $HE_{31}$ even and odd modes is also broken by the nanofiber deformation, which alters the relative phase between the two modes during their propagation along the nanofiber. THG in the $HE_{12}$ mode is then hard to suppress even at the input pump polarization that yields optimum TH-OVG at the output, which can deteriorate the controllability of the third-harmonic topological charge with input pump polarization adjustment. While it is practically difficult to characterize the deformation of nanofiber cross-section, we observe incomplete suppression of THG in the $HE_{12}$ mode in some cases [8], which indicates elliptical deformation of nanofiber cross-section.

The experimental results in Fig. S6 show a typical example of non-ideal TH-OVG with limited topological charge controllability due to the deformation of the nanofiber cross-section, although other key features of TH-OVG are reproduced. The most efficient TH-OVG is observed around a certain pump wavelength (1558 nm in the case of Fig. S6(a)), and a donut-shaped third-harmonic field pattern can be observed with the THG in the $HE_{12}$ mode significantly suppressed (Fig. S6(b)). The cubic pump power dependence of the third-harmonic output power is also obtained, as shown in Fig. S6(c), where a maximum TH-OVG conversion efficiency is measured to be ~$10^{-6}$ at 0.38 W pump power. However, only a single topological charge of either +2 or -2 (+2 in the case of Fig. S6(d)) is revealed via pump polarization control, whereas another with the opposite sign of topological charge is hardly excited with sufficiently suppressing the THG in the undesired $HE_{12}$ mode.

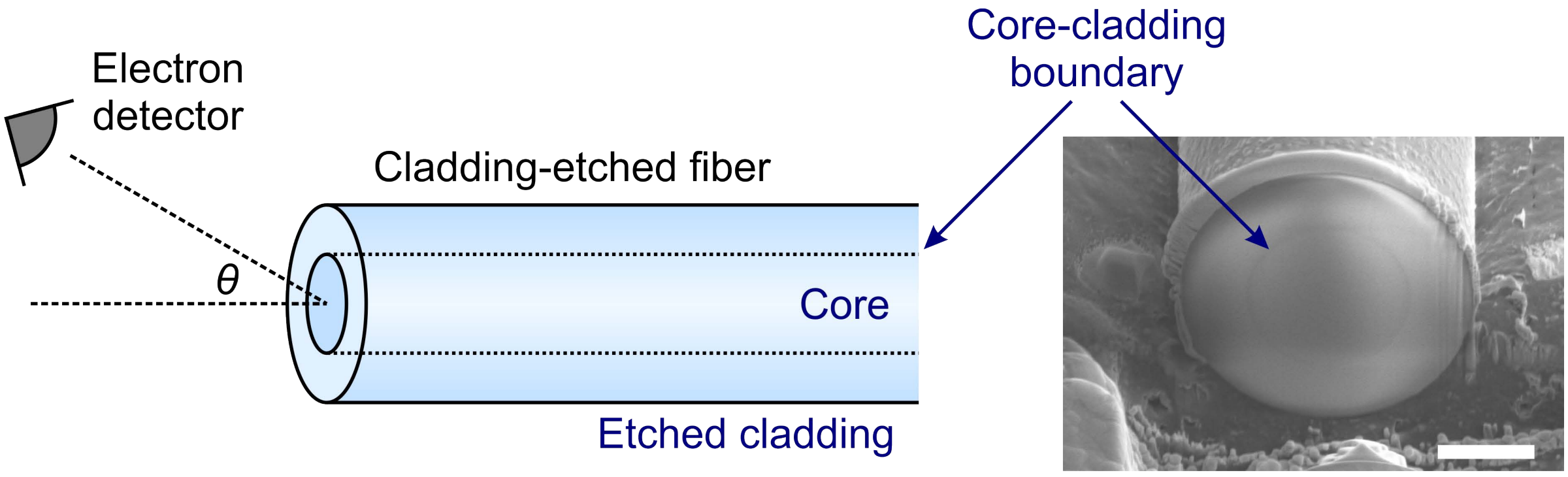

**FIG. S5. Observation of the cross-section of a cladding wet-etched fiber using a focused ion beam (FIB)-scanning electron microscope (SEM) system.** One end of the cladding-etched fiber is polished by FIB milling, and then the end facet is in-situ observed with the SEM. The SEM image on the right displays the end facet, where the slightly darker ellipse is the core-cladding boundary. Note that the cross-section appears generally elliptical on the SEM image because it is detected at an oblique angle $\theta$ from the fiber axis in the FIB-SEM system. The white scale bar in the SEM image corresponds to 5 μm.

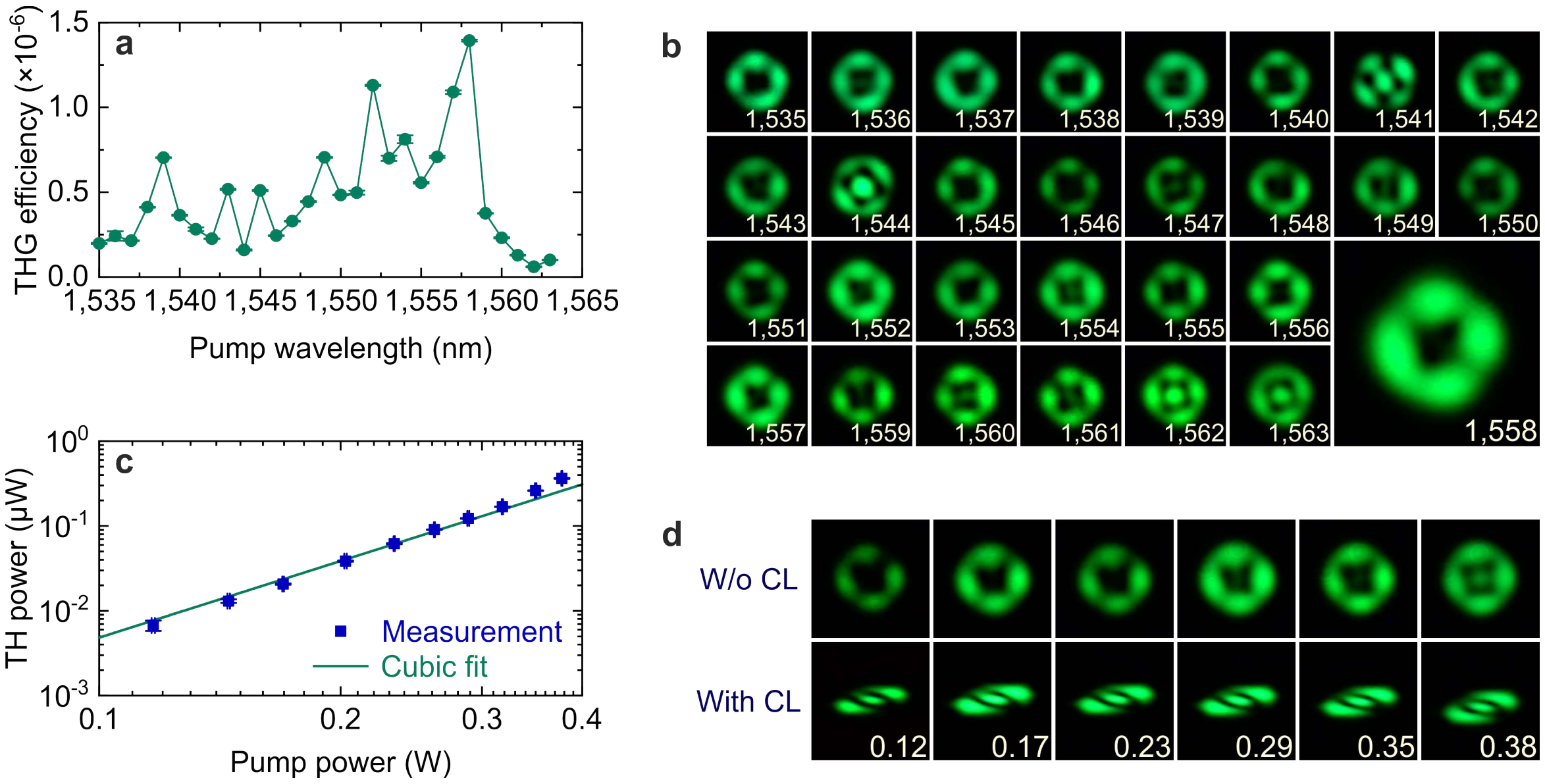

**FIG. S6. Non-ideal third-harmonic optical vortex generation (TH-OVG) with limited topological charge controllability.** (**a**) Measured TH-OVG conversion efficiencies over a pump wavelength range of 1535–1563 nm, while the average pump power is fixed at 0.25 W. The pump polarization state is adjusted to minimize the third harmonic in the undesired $HE_{12}$ mode. (**b**) Far-field profiles of the third-harmonic output signals recorded at different pump wavelengths (white numbers in nanometers). (**c**) Measured third-harmonic (TH) output powers over a range of pump powers at a fixed pump wavelength of around 1558 nm. The turquoise line is a cubic fit to the measurement (blue solid squares). (**d**) Far-field profiles of the third-harmonic output signals without (upper row) and with (lower row) the astigmatic beam transforms via focusing at a cylindrical lens (CL). The numbers in white are the average pump powers in Watt.

## Supplementary References